%% file: main.tex
\documentclass[acmtog]{acmart}
\acmSubmissionID{841}

\usepackage{booktabs} 

\usepackage[ruled]{algorithm2e} 
\usepackage{xcolor}
\usepackage{svg}
\usepackage{ulem}
\usepackage{url}

\SetAlFnt{\small}
\SetAlCapFnt{\small}
\SetAlCapNameFnt{\small}
\SetAlCapHSkip{0pt}

\acmJournal{TOG}
\acmYear{2023} \acmVolume{42} \acmNumber{6} \acmArticle{179} \acmMonth{12} \acmPrice{15.00}\acmDOI{10.1145/3618397}

\setcopyright{acmlicensed}

\begin{document}
\title{Neural Categorical Priors for Physics-Based Character Control}
\begin{teaserfigure}
  \includegraphics[width=\textwidth]{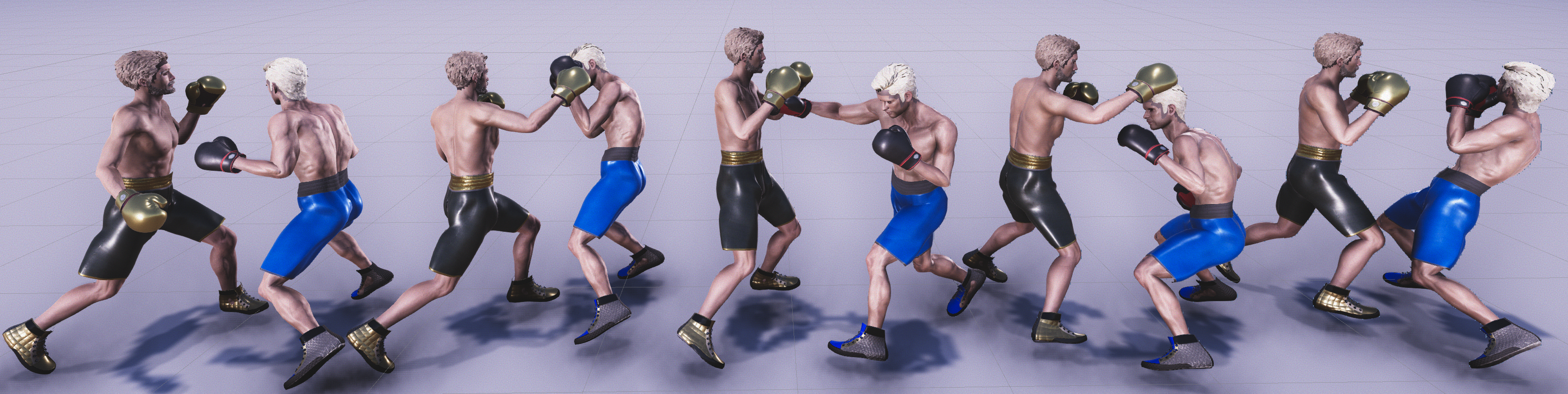}
  \caption{The boxer characters controlled by a neural network trained using our framework can perform professional human boxer strategies (such as defense and dodge) and life-like movements (such as swing, jolt, hook, and bob).}
  \label{fig:teaser}
\end{teaserfigure}

\thanks{$^\ast$Equal contribution. $^\dag$Corresponding author.}
\author{Qingxu Zhu$^\ast$}
\email{qingxuzhu@tencent.com}
\affiliation{%
 \institution{Tencent Robotics X}
 \country{China}}
\author{He Zhang$^\ast$}
\email{herbzhang@tencent.com}
\affiliation{%
 \institution{Tencent Robotics X}
 \country{China}}
\author{Mengting Lan}
\email{carollan@tencent.com}
\affiliation{%
 \institution{Tencent Robotics X}
 \country{China}}
\author{Lei Han$^\dag$}
\email{lxhan@tencent.com}
\affiliation{%
 \institution{Tencent Robotics X}
 \country{China}}

\begin{abstract}
Recent advances in learning reusable motion priors have demonstrated their effectiveness in generating naturalistic behaviors. In this paper, we propose a new learning framework in this paradigm for controlling physics-based characters with improved motion quality and diversity over existing methods. The proposed method uses reinforcement learning (RL) to initially track and imitate life-like movements from unstructured motion clips using the discrete information bottleneck, as adopted in the Vector Quantized Variational AutoEncoder (VQ-VAE). This structure compresses the most relevant information from the motion clips into a compact yet informative latent space, i.e., a discrete space over vector quantized codes. By sampling codes in the space from a trained categorical prior distribution, high-quality life-like behaviors can be generated, similar to the usage of VQ-VAE in computer vision. Although this prior distribution can be trained with the supervision of the encoder's output, it follows the original motion clip distribution in the dataset and could lead to imbalanced behaviors in our setting. To address the issue, we further propose a technique named prior shifting to adjust the prior distribution using curiosity-driven RL. The outcome distribution is demonstrated to offer sufficient behavioral diversity and significantly facilitates upper-level policy learning for downstream tasks. We conduct comprehensive experiments using humanoid characters on two challenging downstream tasks, sword-shield striking and two-player boxing game. Our results demonstrate that the proposed framework 
is capable of controlling the character to perform considerably high-quality movements in terms of behavioral strategies, diversity, and realism. 
Videos, codes, and data are available at \url{https://tencent-roboticsx.github.io/NCP/}.
\end{abstract}
%
%


%
%


\maketitle
\input{body}

\end{document}

%% file: body.tex

\section{Introduction}
Recent breakthroughs in computer vision and natural language processing 
reveal that pre-trained deep representations from massive datasets are extremely expressive to maintain comprehensive knowledge in these datasets. 
The pre-trained models enable the possibility of fast adaptation and learning to solve complex downstream tasks.
Inspired by this,
many control models for physically simulated characters are proposed to
initially learn latent representations of motion clips;
once the latent representations are shaped, these prior models are easily reused to initialize or guide an upper-level control model and facilitate its training in new downstream tasks.
For example,~\cite{merel2020catch, liu2022motor,yao2022controlvae, won2022physics} employ $\beta$-VAE~\cite{higgins2017beta}, 
a variant of a variational autoencoder (VAE)~\cite{kingma2013auto},
to leverage between the reconstruction performance of motion clips and the constraint on the learned latent distribution, which is commonly assumed as a Gaussian. 
Alternatively, another branch of studies embraces the generative adversarial network (GAN)~\cite{goodfellow2020generative} that employs a discriminator to help the control model generate similar movements hard to be distinguished from the motion clip data distribution.
For example,
\citep{peng2022ase,juravsky2022padl} apply generative adversarial imitation learning~(GAIL)~\cite{ho2016generative} 
to learn a control policy imitating expert demonstrations. 
Unfortunately, generative adversarial methods are notorious for unstable training and insufficient diversity of generated samples. 
\cite{peng2022ase} encounters a similar situation and has to employ an additional regularization term on the latent variable to force diverse skills. 
The most recent approach~\cite{tessler2023calm} adopts a conditional discriminator, forcing the control policy to perform distinct and diverse behaviors stick to a given latent variable. 
However, generating diverse movements sufficiently covering the entire data distribution is still a critical issue.

Overall, maintaining both realism and diversity is challenging for physics-based character control. 
In this paper, we consider this problem from the perspective of investigating the expressive ability of the latent space. 
Both existing VAE- and GAN-based methods commonly assume Gaussian or hypersphere latent spaces. 
Under the assumption, the control model naturally faces the trade-off between the reconstruction performance and the compatibility with the prior distribution. 
For example, in $\beta$-VAE based methods, if we put loose constraint over the KL-divergence between the learned representation distribution and the Gaussian, i.e., $\beta$ is small, the reconstruction loss could be lower while the learned latent space might stay far from Gaussian, preventing 
efficient
sampling from this prior.
On the contrary, if we penalize the KL-divergence over aggressively, i.e., $\beta$ is large, the reconstruction performance will be deteriorated too severely to produce realistic behaviors.
GAN-based methods usually suffer from mode collapse and limited diversity that partially matches the original data distribution.
A practical way to alleviate this issue is to leverage a more expressive prior distribution for the latent representations.

The discrete information bottleneck, introduced in Vector Quantized Variational Autoencoder (VQ-VAE)~\cite{van2017neural, razavi2019generating}, 
has shown promise in generating diverse and high-fidelity images using discrete latent representations.
Theoretical results for VQ-VAE~\cite{roy2018theory} have demonstrated that the distribution of latent variable equals to a special case of mixture-of-Gaussians with 
nearly zero covariance and uniform prior over cluster probabilities. VQ-VAE maintains a codebook constituted with a certain number of vector quantized codes to represent the latent space. 
Once the VQ-VAE has completed training, 
we need to additionally learn a categorical prior distribution over the codes, supervised by the outputs of the encoder given the data. 
The learned representation in the codebook together with the prior distribution are used to generate new random samples.

In this paper, we
propose a conditional discrete information bottleneck structure in the control policy to generate motor primitive actions. 
The control policy is trained to imitate movements from unstructured motion clips using Reinforcement Learning (RL), conditioned on the current character state.
The vector quantized codes learned at this stage offer a discrete latent representation for the motion clips.
At the next stage, we remove the encoder and train a categorical prior distribution without conditioning on the motion clips to match the encoder's output given the motion data. 
Since the trained prior distribution is categorical and represented as a neural network, we refer to it as the Neural Categorical Priors (NCP). 
Then, by drawing samples from NCP, the control policy could output actions with life-like behaviors thanks to the expressive ability of vector quantized codes.
A well-trained prior distribution can accurately reveal the original data distribution given sufficient data. 
While this is not an issue for computer vision and natural language processing problems because the datasets in these fields are sufficiently large with billions of samples, in our case we only have tens of public motion clips. 
For small datasets, a critical issue is the data imbalance. 
For example, some rare yet fancy movements, like combination of punches in boxing motions, will be rarely generated by sampling from the trained prior distribution.
This heavily lowers the exploration efficiency when reusing the prior distribution for upper-level policy in solving downstream tasks.
None of the existing methods resolve the issue of diversity from the perspective of imbalanced motion priors.
Therefore, instead of directly reusing the trained prior distribution, we propose a new technique named prior shifting to adjust the prior distribution. 
Specifically, we take advantage of the spirit of curiosity-driven RL 
by introducing some curiosity-driven rewards to encourage the shifted distribution evenly cover all possible motion clips. 
Finally, we build upon an upper-level policy with discrete action space on top of the fixed decoder trained at the imitation learning stage and reuse the shifted NCP via a KL-regularized term between the upper-level policy and the shifted NCP. The entire structure is trained using RL to optimizing the downstream task reward.
It is worth mentioning that the NCP from the discrete information bottleneck enables the chance to adjust the prior distribution, while the case turns to be less flexible for Gaussian priors. Moreover, using NCP, the upper-level policy naturally outputs discrete actions to choose from the vector quantized codes, whose cardinality usually ranges from tens to hundreds. This creates a very small and compact exploration space for the upper-level policy, compared to exploring in a high dimensional continuous space as used in VAE- and GAN- based methods.

In the experiments, we conduct comprehensive studies on two challenging benchmark datasets with motion clips of sword \& shield~\cite{peng2022ase} and boxing sports~\cite{won2021control}. We compare our method with the state-of-the-art methods, including $\beta$-VAE based method
and GAN-based methods~\cite{peng2022ase,tessler2023calm}. 
We first compare these methods by evaluating the randomly generated movements from their learned motion priors. 
Quantitative metrics on realism and diversity demonstrate that our method 
outperforms others with a large margin.
Moreover, we experiment in solving two downstream tasks following~\cite{peng2022ase,won2021control}, including sword \& shield strike and two-player boxing match. 
For the later task, we employ prioritized fictitious self-play (PFSP)~\cite{vinyals2019grandmaster} and the convergent policy shows surprisingly close attacking and defensing strategies compared to real two-player boxing sports. 

In summary, our main contributions are that 1) we introduce conditional discrete information bottleneck to the control policy for physics-based characters, where the learned discrete representations benefit from a more expressive latent space; 2) we propose prior shifting to balance the learned prior distribution to cover all possible motion clips, substantially enhancing the diversity of generated movements; 3) quantitative evaluations on sword \& shield and boxing sports benchmarks show that our method produces considerably improved behavioral realism and diversity over compared methods.

\section{RELATED WORK}

\subsection{Physics-based Motion Controller}
The development of control systems for physically simulated characters has been a longstanding topic in the field of computer animation.
Early works frequently employed heuristic-based algorithms to generate dynamic behavior for the animated characters \cite{raibert1991animation, hodgins1995animating}. However, these methods typically necessitated unique designs for individual behaviors.
To produce control policies that are more general and resilient, optimization-based approaches, such as trajectory optimization and reinforcement learning, have been extensively investigated \cite{mordatch2012discovery, tan2014learning, yin2008continuation, macchietto2009momentum, xie2020allsteps, yin2021discovering, yu2018learning, peng2016terrain, geijtenbeek2013flexible, brown2013control, da2017tunable}.
While these models can achieve robust control, the task of devising an objective for attaining life-like motions is implicit and intricate. 

To enhance motion realism,
physics-based methods frequently involve the 
usage of motion capture data \cite{da2008simulation, jain2011modal, zordan2002motion, kwon2017momentum, ding2015learning, muico2009contact}.
One direct way to integrate the motion reference data into a physics-based controller is 
motion tracking, where the objective is to minimize the pose error between the simulated characters and the target motion references \cite{yin2007simbicon, sok2007simulating, lee2010data, liu2016guided, liu2010sampling, muico2011composite}.
Based on the development of deep reinforcement learning, highly dynamic motions such as backflipping and spinning kick can be 
accurately tracked by deep neural networks and applied in different scenarios \cite{peng2018deepmimic, won2019learning, lee2022deep, lee2021learning}. 
However, scaling up these techniques to handle large datasets and complex scenarios still remains challenging.
An intuitive approach to address this issue is to employ the state machine \cite{liu2018learning} or mixture-of-experts (MOEs) \cite{peng2019mcp} 
to integrate diverse pre-trained policies, yet these structures 
require preservation of all sub-policies and careful management of transitions 
among them.
Alternatively, some studies opt to directly track a kinematics-based controller, which can effectively regulate motion transitions \cite{bergamin2019drecon, park2019learning, won2020scalable}.
Along this direction, recent work has applied world model-based techniques to simultaneously track the kinematics controller and estimate a dynamic model, resulting in significant improvement in training speed \cite{fussell2021supertrack}.
Nonetheless, the capability of these models is restricted by the kinematics controller, and creating an effective kinematics controller for tasks such as multi-character interactions 
remains challenging.

Similar to these tracking-based methods, 
our work utilizes an imitation learning 
stage
to build the lower-level controller. The difference is that we employ a single generative model to track all the motion data while learning compact discrete latent representations that can be reused for complex downstream tasks, such as two-player boxing.

\subsection{Deep Generative Model for Motion Control}
The advancement of deep learning has catalyzed the investigation and utilization of generative models. 
For kinematics-based motion controls, VAE \cite{ling2020character}, GAN \cite{li2022ganimator}, and flow-based models \cite{henter2020moglow} have been thoroughly investigated.
While kinematics-based motion controls can synthesize high-quality animation, they primarily focus on locomotion tasks. 
For more complex scenarios, such as interacting with objects and other characters, 
reference data are highly demanded.
Motion Recommendation \cite{cho2021motion} utilized a combination of a finite state machine and the VQ-VAE model to construct a compact motion graph that can handle a diverse range of motions.
However, this method only utilizes the discretization space from the VQ-VAE encoder for motion clustering, and a motion matching technique is still necessary for intricate interaction tasks.
Although our work shares a similar concept of using the VQ-VAE model, we focus on the expressive power of the discrete latent sapce and train an end-to-end control policy absorbing the information bottleneck structure instead of discretizing the data space.

Recently, an increasing number of studies have utilized generative models to improve physical simulation control and implement them in complex interactive environments. Specifically, the GAN-based model and VAE-based model have been primarily investigated. 
Inspired by generative adversarial imitation learning (GAIL)~\cite{ho2016generative}, adversarial motion priors (AMP)~\cite{peng2021amp} have been proposed as the objective for motion imitating. Such an adversarial objective can provide controllers with the ability to compose appropriate reference motions and even generate novel transitions to complete downstream tasks. 
Despite its ability to generate high-quality results on complex tasks like character-scene interaction \cite{hassan2023synthesizing}, this model requires re-training from scratch each time the task changes. Furthermore, achieving balance in adversarial training is not trivial.
To improve the model reusability, pre-trained motion priors (ASE)~\cite{peng2022ase} based on the GAIL structure is further introduced, with which a diverse set of new tasks can be effectively learned \cite{juravsky2022padl}.
However, mode-collapse is an inherent problem of the GAN model, although techniques such as conditional adversarial imitation learning can alleviate this issue \cite{tessler2023calm}, 
it is still challenging to cover all the motion data in the latent space. 
Another potent generative model is the VAE-based model, where the distillation technique has been utilized to combine different pre-trained policies. The outcome policy can then act as the low-level controller for downstream tasks such as carrying and catching \cite{merel2020catch}. 
By combining with world model-based methods, the low-level controller can be learned and reused efficiently for higher-level training~\cite{yao2022controlvae, won2022physics}. 
Although the traditional VAE-based model has the potential to compress all motions into the latent space, achieving a balance between the reconstruction quality and a low KL divergence can be challenging. A loose constraint over KL-divergence can result in a non-Gaussian latent space, while a higher weight for the KL loss often leads to poor motion quality. 

The VQ-VAE model applied in our study belongs to the VAE-based family. However, in comparison to previous works, we demonstrate that our model can achieve high-quality motion tracking while generating diverse motions to accomplish complex tasks such as two-character competition. 

\subsection{Multi-player Animation}

A classic approach to synthesizing large-scale character animation is motion patches \cite{lee2006motion}, where each patch is a building block annotated with specific motion data in the scene. 
By precomputing the interaction information between multiple characters in each building block, patch-based methods can be leveraged for multi-character animation \cite{shum2008interaction, kim2012tiling, won2014generating, kim2009synchronized}.
In terms of contact-rich and close interaction between multiple characters, the interaction mesh representation has been exploited to preserve the spatial information between body parts and objects \cite{ho2010spatial}. 
However, capturing the motions of multiple avatars simultaneously is a difficult task that often requires extensive post-processing.  
As an alternative, some approaches utilize single-character motion data by learning competition or collaboration policies.
Building on the motion graph structure, a transition of each individual player can be planned by dynamic programming, dynamic Bayesian network, and game theory \cite{lee2004precomputing, kwon2008two, wampler2010character}.
Using an action-level graph, a game tree structure can also be established to organize collaborative and adversarial behaviors among multiple characters \cite{shum2007simulating, shum2008simulating, shum2010simulating}.
When data is limited, dynamic constraints and physics-based spacetime optimizations are often utilized to assist motion exploration while maintaining the realism of the character \cite{liu2006composition}.

Recent progress has been made in this direction by combining the physics-based simulation with deep reinforcement learning.
A two-stage framework was developed to control two-player competitive sports that involve physical interaction and high degrees-of-freedom joint angle spaces \cite{won2021control}. In the first stage, the basic skills are learned by the individual character with an encoder-decoder structure and imitation learning. The mixture-of-experts decoder is then incorporated to combine skills from different imitation policies. Our framework is closely related to this type of hierarchical structure. However, instead of blending the latent variables from separate experts, our high-level controller directly leverages the latent embedding from a single generative model, which enables us to generate more continuous motions. Additionally, it is easy to integrate a learned motion prior in our model to ensure both the motion quality and the diversity in high-level competition tasks.

\section{BACKGROUND}
\subsection{Reinforcement Learning}
Reinforcement learning solves an online decision problem, in which an agent interacts with an environment as a Markov Decision Process (MDP). 
At time step $t$, the agent performs an action $a_t$ conditioned on a state $s_t$, and then receives a reward $r_t$ and next state $s_{t+1}$ from the environment.
The probability of state transition can be described as $p\left(s_{t+1} | s_t, a_t\right)$. The objective of the agent is to learn a policy $ \pi(a_t | s_t)$ that maximizes the expected cumulative reward $G(\tau)$ over trajectories
\begin{equation*}
    \mathcal{J}_{RL}(\pi)=\mathbb{E}_{\tau \sim \pi(\tau)}[G(\tau)]=\mathbb{E}_{\tau \sim \pi(\tau)}\left[\sum_{t=0} \gamma^t r_{t}\right],
\end{equation*}
where $\pi(\tau)=p\left(s_0\right) \prod_{t=0} \pi\left(a_t | s_t\right) p\left(s_{t+1} | s_t, a_t\right)$ 
denotes the probability of a trajectory $\tau$; $p\left(s_0\right)$ is some distribution of the initial state; 
and $\gamma \in [0, 1]$ is a discount factor. 
For optimizing the RL problem, we choose the PPO~\cite{schulman2017proximal} 
algorithm throughout this paper.

\subsection{Vector Quantized Variational Autoencoder}
Vector Quantized Variational Autoencoder (VQ-VAE)~\cite{van2017neural, razavi2019generating} is a deep representation learning method that combines the strengths of Variational Autoencoders (VAEs) and Vector Quantization (VQ) to learn a discrete codebook with a set of representative latent embeddings. 
In VQ-VAE, the encoder maps the input to one latent embedding (code) in the codebook. 
Then, the decoder maps the code 
into an output to recover the input. 
The VQ-VAE training loss consists of a reconstruction loss and a commitment loss as denoted below
\begin{equation*}
    \mathcal{L}_{\text{VQ-VAE}}=-\log p\left(x | z^q(x)\right)+\left\|\operatorname{sg}\left[z^e(x)\right]-e\right\|_2^2+\beta\left\|z^e(x)-\operatorname{sg}[e]\right\|_2^2,
\end{equation*}
where $-\log p\left(x |z^q(x)\right)$ is the reconstruction loss for a data sample $x$ and $z^q(x)$ is the quantized vector; $z^e(x)$ is the output of the encoder and $e$ is the nearest code to $z^e(x)$;
$\operatorname{sg}$ indicates a stopgradient operator;
$\beta$ is a hyperparameter to balance the last two terms in the commitment loss.
The reconstruction loss measures the difference between the input data and the reconstructed output, while the commitment loss encourages the encoder to commit to a discrete code by penalizing the distance between the encoding and its nearest embedding in the codebook.
Additionally, a prior should be introduced to control the distribution of embeddings and encourage diversity in the generated samples. 
When training the VQ-VAE, the prior is kept constant and uniform. 
After its training, a 
parameterized
categorical prior distribution is fit to match the distribution of these discrete codes using maximum likelihood estimation.
The decoder network can then generate high-quality samples by sampling codes from the learned categorical prior.

\begin{figure*}[h]
  \centering
  \includegraphics[width=\linewidth]{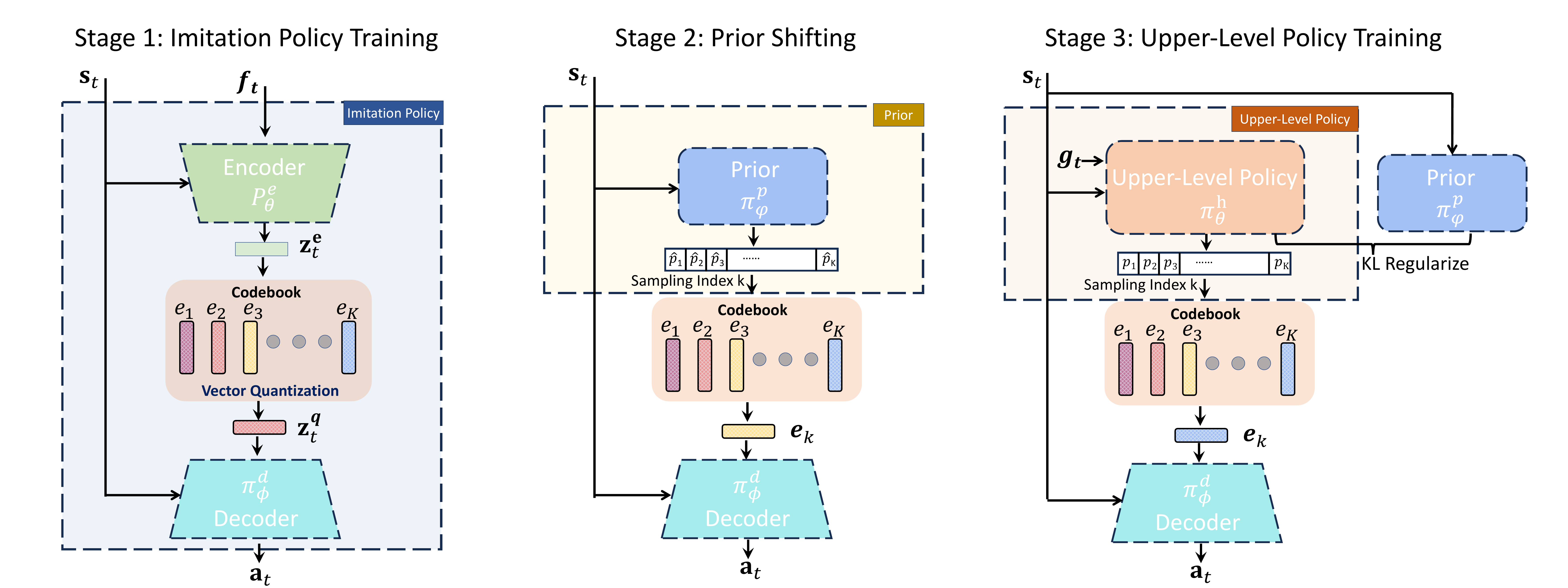}
  \vskip -0.1in
  \caption{The Framework overview of NCP consists of three stages of training. The first stage involves imitating lifelike movements from unstructured motion clips. In the second stage, a neural categorical prior is trained using prior distillation and adapted using prior shifting to balance different movements. In the final stage, the upper-level policy is trained under the KL-divergence constraint to remain close to the previously trained categorical prior.}
  \label{fig:overview}
\end{figure*}

Overall, VQ-VAE provides a powerful framework for learning discrete representations of data, which can be used for a variety of tasks, including compression, generation, and clustering.

\section{OVERVIEW}
Fig.~\ref{fig:overview} provides an overview of the proposed NCP framework, which is composed of three stages of training described in the rectangles. 
In the first stage, we use deep RL to initially track and imitate life-like movements from unstructured motion clips using the conditional discrete information bottleneck structure, adapted from that of VQ-VAE. 
In the second stage, we utilize a prior distillation method to train a neural categorical prior, and then adapt it using a technique named prior shifting to flatten the distribution over distinct states. The shifted categorical prior network is later used for rolling out random trajectories and regularizing the upper-level controllers. 
At the final stage, we train the upper-level policy under the KL-divergence constraint to let it stay close to the previously trained categorical prior. In the following subsections, we will detail each of these stages.

\section{The IMITATION POLICY}
Most existing methods 
of learning reusable motion priors adopt an encoder-decoder architecture or adversarial generative fashion. 
Both VAE- and GAN-based models have been widely investigated for training an imitation policy given expert demonstrations. 
As introduced previously, each category of these approaches
bears certain limitations.
Since the VQ-VAE model belongs to the VAE families, in this section, we first discuss VAE-based methods in detail. 
Then, we propose the control policy using the discrete information bottleneck to imitate motion clips.

\subsection{Limitations of VAE-based Methods}

For VAE-based models, one common choice for the prior distribution $p(z)$ is the standard multivariate normal distribution $\mathcal{N}(0, I)$. For instance, \cite{yao2022controlvae, won2022physics} employ a conditional $\beta$-VAE \cite{higgins2017beta} to imitate motion skills, where the prior is formulated as a conditional distribution $p\left(z_t | s_t\right) \sim \mathcal{N}(0, I)$. The following loss function is used to train the conditional VAE
\begin{equation}\label{vae-objective}
	\mathcal{L}_{\beta\text{-VAE}}=\mathcal{L}_{\text {rec}} + \beta \cdot D_{K L}\left(\mathcal{N}\left(\mu_t, \sigma_t\right) \| \mathcal{N}(0, I)\right)
\end{equation}
where $\mathcal{L}_{\text {rec}}$ is the reconstruction loss, which measures the difference between the generated character state and target state in motion clips; 
$D_{KL}(\cdot \| \cdot)$ measures the KL-divergence of the prior distribution $\mathcal{N}(0, I)$ and posterior distribution $\mathcal{N}\left(\mu_t, \sigma_t\right)$ with $\mu_t$ and $\sigma_t$ being estimated. 
The parameter $\beta$ controls the strength of the regularization term and trades off between reconstruction accuracy and latent space regularization.
The information from the motion clips is compressed into the latent variable $z$, forming an information bottleneck structure.

However, it is difficult to preserve both sample quality and distribution constraint through a compact latent space using 
$\beta$-VAE.
That is, if $\beta$ is too large, excessive penalty on the KL-divergence could lead to posterior collapse, i.e., the KL-vanishing problem~\cite{shao2020controlvae}.
On the opposite, loosing the KL-divergence could enhance the
reconstruction performance, but it might cause the posterior distribution to deviate from the prior distribution, making it challenging to draw samples from the prior distribution.

\subsection{Conditional Discrete Information Bottleneck}
In this section, we introduce the discrete information bottleneck into an imitation policy.
Our imitation policy contains a conditional encoder-decoder structure. The encoder $p^{e}\left(z_t^e | s_t,f_t\right)$ takes input the state $s_t$ and the future trajectory $f_t$ from motion clips, and map them to a latent variable $z_t^e$. 
The variable $z_t^e$ will be quantized to an embedding $z_t^q$ from a finite number of trainable embeddings, which are referred to as codes. 
The decoder $\pi^{d}(a_t | s_t,z_t^q)$ takes input the state $s_t$ and the code $z_t^q$, and outputs $a_t$ as the motor primitive action, i.e., the target joint positions for PD controllers.

The trainable latent embedding codebook is represented as $\mathbf{e}\in R^{K \times D}$, where $K$ is the size of the discrete latent space or the number of codes, and $D$ is the dimensionality of each latent embedding vector or code $e_i$, resulting in $K$ embedding vectors $e_i \in R^D, i \in 1,2, \ldots, K$. 
Given the codebook $\mathbf{e} \in R^{K \times D}$, the encoder output $z_t^e$ is mapped to the nearest code in $\mathbf{e}$. That is, the discrete latent variables $z_t^q$ is calculated through the nearest neighbor look-up as denoted below
\begin{equation}\label{vq-vae-quantizied}
z_t^q=e_k, \text{ where } k=\arg\min_{j}\left\|z_t^e-e_j\right\|_2.
\end{equation}

In other words, the posterior categorical distribution $q(\mathbb{I}(z^e))$ of the embedding vector index is defined as follows
\begin{equation}\label{vq-vae-posterior}
	q(\mathbb{I}(z^e) =k)= \begin{cases}1 & \text { if } k=\underset{j}{\arg\min}\left\|z^e-e_j\right\|_2, \\ 0 & \text { otherwise. }\end{cases}
\end{equation}
where $\mathbb{I}(z^e)$ is an indicator function returning the index of 
the selected code for encoding $z^e$.

\subsection{Latent Space Analysis}
We discuss the benefits of using discrete information bottleneck. As revealed in~\cite{roy2018theory, bishop2006pattern}, the quantization process of VQ-VAE, the K-means, and the EM algorithm share many similarities. Considering a Gaussian mixture model with means $\mathbf{e}$ and covariance matrices $\epsilon \mathbf{I}$ of the mixture components, where $\epsilon$ is a shared variance parameter for all of the components. 
Give a particular data point $z^e$, the posterior probability of $\mathbb{I}(z^e)$ can be formulated using the EM algorithm as follows
\begin{equation}
q(\mathbb{I}(z^e)=k)=\frac{\delta_k \exp \left\{-\left\|z^e-e_k\right\|^2 / 2 \epsilon\right\}}{\sum_j \delta_j \exp \left\{-\left\|z^e-e_j\right\|^2 / 2 \epsilon\right\}},
\end{equation}
where $\delta_k$ is the prior probability of the $k$-th component of the mixture. 
If we consider the limit $\epsilon\rightarrow0$, as shown in~\cite{bishop2006pattern}, the term 
$\delta_j \exp \left\{-\left\|z^e-e_j\right\|^2 / 2 \epsilon\right\}$ for which $\|z^e-e_j\|^2$ is smallest will go to zero most slowly. Therefore, all terms in the posterior distribution approach zero except for one term, which approaches 1. 
This reveals in this extreme setting, the data point $z^e$ is assigned to a cluster in a hard manner, demonstrating the same process as described in Eq.~\eqref{vq-vae-posterior}. 
So, the latent distribution of the discrete information bottleneck is actually a multivariate Gaussian mixture model with covariance matrices $\epsilon \mathbf{I}$, where $\epsilon$ is close to 0. 
This provides theoretical evidence to support that the discrete information bottleneck is likely to achieve superior performance when utilizing finite discrete latent variables.

On the other hand, the discrete bottleneck can lead to a compact latent space. In contrast to the original VQ-VAE, which 
selects a set of embeddings for the decoder (where each embedding corresponds to a pixel of image in their case), our framework utilizes only a single embedding that is passed to the decoder.
Therefore, the discrete information bottleneck yields a compressed yet interpretable latent space consisting of $K$ discrete points in continuous space, as opposed to the continuous latent spaces of VAE- and GAN-based methods. 
The usage of discrete spaces can offer significant advantages for downstream tasks. In particular, exploration in discrete space could be much easier compared to exploration in continuous space, as discussed in several previous works~\cite{tang2020discretizing, andrychowicz2020learning}.

Moreover, VQ-VAE has the advantage of applying a constant KL-divergence between the prior and posterior distributions, avoiding the need to balance the reconstruction loss and KL-divergence as considered in previous $\beta$-VAE based works. This property makes the training process simpler and more stable. 

\subsection{Training}
To enable a single control policy to imitate all motion clips, our framework employs a tracking-based way. Unlike~\cite{yao2022controlvae, won2022physics}, who train a world model to construct a complete VAE to reconstruct the states, we follow the tracking method proposed in~\cite{peng2018deepmimic}, which defines some imitation rewards and use RL to train the model. 
In our case, the reward $r_t$ is defined as
\begin{equation}
r_t=\left\|\mathbf{W}^{1 / 2}\Omega(s_{t+1} - s_{t+1}^*)\right\|_2^2,
\end{equation}
where $\mathbf{W}$ is a diagonal weight matrix selected based on empirical evidence to balance the magnitude of each imitation reward; 
$\Omega(s)$ is a function specifically designed to determine the reward associated with a given state difference. The subsequent target state in motion clips is represented as $s_{t+1}^*$, and state $s_{t+1}$ represents the actual next arrived state calculated by the simulator. Using RL, we can accurately imitate of a broad range of motion clips.

Similar to Eq.~\eqref{vae-objective}, the loss in our framework also composes of two parts, an RL objective to imitate the movements in motion datasets and a commitment loss to learn a powerful latent representation. 
The overall objective is formulated as
\begin{equation}
	\begin{gathered}
\underset{\theta,\phi,e}{\text{maximize}} 
\ 
\mathbb{E}_{p_{\theta}^{e},\pi_{\phi}^{d}} 
\bigg[ \sum_{t=0}^{T-1} \Big( \gamma^t r_t
-\left\|\operatorname{sg}\left[z_t^e\right]-e\right\|_2^2
-\beta\left\|z_t^e-\operatorname{sg}[e]\right\|_2^2 \Big)\bigg]
\end{gathered}
\end{equation}
where $\theta$ and $\phi$ are the parameters of the encoder $p^e$ and decoder $\pi^d$, respectively;
$\operatorname{sg}$ indicates the stopgradient operator, which is defined as an identity during forward computation and has partial derivatives that equal to zero; 
the notation $e$ without any indexing indicates the nearest neighbor look-up from the codes; according to~\cite{van2017neural, razavi2019generating}, $\beta$ is a quite robust parameter in balancing the two terms in the commitment loss.

\subsection{Prioritized Sampling}
\label{sec:prioritized_sampling}
We propose to imitate all motion clips into a single policy. 
Similar to the data imbalance issue commonly encountered in standard supervised learning, 
some motion clips that are rare from the dataset might be underfitting in the imitation policy. 
For example, some agile movements like the combination of punches are captured less frequently compared to common locomotion behaviors like walking. 
Worse still, due to the dynamics, these agile movements are usually more challenging to imitate than other motion clips. 
The consequence is that the imitation policy fails to reproduce such movements.
Instead of uniformly sampling motion clips in the dataset, we utilize prioritized sampling, where the motion clip $m_i$ from the dataset $\mathcal{M}$ is sampled with probability
\begin{equation} 
p_i=\frac{f(R_{m_i})}{\sum_{m \in \mathcal{M}} f(R_m)},
\end{equation}
where $f:[0,1] \rightarrow[0, \infty)$ is a weighting function and $R_m\in[0,1]$ represents the normalized rewards obtained in the RL tracking task for motion clip $m$.

By choosing $f(x)=(1-x)^{\alpha_1}$, the policy keeps attention on these challenging motion clips. The sharpness of the distribution can be determined by choosing $\alpha \in \mathbb{R}_{+}$. 
Similar practices have been reported in~\cite{xie2022learning, won2019learning}.

\section{PRIOR}
When reusing the pre-trained representation in downstream tasks, existing VAE-based methods often discard the trained encoder in the previous stage, and directly create a upper-level policy to explore the latent space to drive the fixed decoder.
In fact, the encoder contains valuable information from the original data distribution.
In this section, we explain how to make use of the encoder $p^{e}(z_t^e | s_t,f_t)$ learned in the imitation stage to learn a prior distribution. This prior can then be used to generate a variety of motions and facilitate downstream task learning.

\subsection{Prior Distillation}
Similar to VQ-VAE, a constant and uniform prior is kept during the training of the imitation policy. 
After that, we additionally train a categorical prior network $\pi_{\varphi}^{p}(\cdot | s_t)$ with parameter $\varphi$ to fit the categorical distribution over the codes given only $s_t$,
without knowing the future trajectory $f_t$. 
We formulate this process as a policy distillation problem~\cite{rusu2015policy}. 
That is, we fit the distribution of the encoder's output $z^e$ using $\pi_{\varphi}^{p}(\cdot | s_t)$ with parameter $\varphi$. 
In this fitting process, we collect trajectories using the current $\pi_{\varphi}^{p}(\cdot | s_t)$, and the trajectories are used to optimize the following distillation objective
\begin{equation}\label{kl_loss}
    \underset{\varphi}{\text{minimize}}\ 
    \mathbb{E}_{\tau \sim \pi_{\varphi}^{p}, \atop f_t\sim\mathcal{M}}\ 
    \sum_{t=0} \mathrm{KL}\left(p_{\theta}^e(\cdot | s_t,f_t) \| \pi_{\varphi}^{p}(\cdot| s_t)\right),
\end{equation}
where $\tau$ is the trajectory generated by the prior network $\pi_{\varphi}^p$ and $f_t$ is some future trajectory randomly sampled from the motion dataset $\mathcal{M}$. 
This process is repeated iteratively until convergence of the prior network.
The trained prior networks can be used to generate random naturalistic movements following the original data distribution.
More explanations for expanding Eq.~\eqref{kl_loss} can be found in the Appendix \ref{appendix:explanation}.

\subsection{Prior Shifting}
After the prior distillation process, sampling a sequence of codes from the distribution $\pi_{\varphi}^{p}(\cdot | s_t)$ allows the decoder to execute high-quality movements. However, due to the data imbalance of $\mathcal{M}$ in the prior distillation process, sampling codes from the trained prior distribution naturally drive the decoder to perform movements that appeared more frequently in the data.
This could raise diversity issues when the data is imbalanced.

To facilitate the exploration in downstream task learning with unknown skill preference, it is desired to have a balanced prior capable of performing a diverse range of movements with nearly uniform probability over distinct motions in the dataset. 
Inspired by the curiosity-driven exploration methods \cite{strehl2008analysis, ostrovski2017count, bellemare2016unifying}, we propose a count-based RL method to fine-tune the prior distribution.

In prior shifting, the prior network is initialized from the distillation stage and the decoder is fixed. The latent variables sampled from the prior drive the decoder to generate movements. To adjust the prior distribution, we employ a heuristic count-based reward defined as
\begin{equation}
    r(s)=\sqrt{\frac{N_k}{\hat{N}(s)}}
\end{equation}
where $\hat{N}(s)$ is a pseudo-count of a continuous state $s$, and $N_k$ is a constant used to scale the magnitude of the reward.
To estimate the pseudo-count $\hat{N}(s)$ in continuous space, density models have been applied in existing works such as~\cite{ostrovski2017count, bellemare2016unifying}. In our case, considering the limited number of frames in motion clips, we estimate the pseudo-count by simply using a motion matching method without the need for precise density model fitting. To achieve this, a metric is defined to select motion frames in motion clips that are close to the sampled data within a given threshold, and the count of every matched frame is increased by 1 and stored in a table with a size being equal to the number of frames. To stabilize the training, only the latest $T_k$ generated states are counted, where $T_k$ is a constant. The reward is then calculated using the count of the frame with the highest matching score. In cases where a state is unable to match any frame in the dataset given the threshold, we assign a zero reward to penalize unnatural behaviors.


\section{UPPER-LEVEL POLICY}
Now, we are ready to solve downstream tasks. The pre-trained decoder $\pi_{\phi}^d$ and shifted prior distribution $\pi_{\varphi}^p$ have absorbed sufficient knowledge on both movement quality and diversity. Reusing these pre-trained networks,
the remaining job is simply building an upper-level policy $\pi_{\vartheta}^h(\cdot| s_t,g_t)$ with parameter $\vartheta$ for choosing a discrete code to drive the fixed decoder for producing actions, where $g_t$ indicates some task-specific goals that are only relevant to the downstream tasks.
Below, we introduce how to reuse the shifted prior $\pi_{\varphi}^p$ for training the upper-level policy $\pi_{\vartheta}^h$.

\subsection{KL-Regularized Training}

Specifically, in addition to the RL loss that maximizes the cumulative task reward, 
we apply a KL-regularized
term to ensure that the upper-level policy $\pi_{\vartheta}^h$ remains close to the pre-trained prior distribution $\pi_{\varphi}^p$. Moreover, an entropy bonus term is also incorporated to promote exploration and prevent  convergence to sub-optimal solutions. 
The loss function is formulated as follows
\begin{equation}
\begin{split}\label{eq:hlc}
    \text{maximize}_{\vartheta}\ \mathbb{E}_{\pi_{\vartheta}^h}\left[
    \sum_{t=0}\gamma^t r_t\right] & -\alpha_{\mathrm{KL}}\cdot \mathrm{KL}\left(\pi_{\vartheta}^h(\cdot| s_t,g_t)|| \pi_{\varphi}^p(\cdot |s_t) \right) \\
    &+ \alpha_{\mathrm{H}}\cdot \mathrm{H}(\pi_{\vartheta}^h),
\end{split}
\end{equation}
where $r_t$ here indicates the downstream task reward and $\mathrm{H}(\pi_{\vartheta}^h)$ is the entropy.
This objective aims to maximize rewards while letting the upper-level policy stay close to the learned prior to performing movements in motion clips. 
Therefore, it encourages exploration and prevents the policy from converging to a deterministic solution, which promotes the learning of multiple solutions to a task to enhance diversity.

\subsection{Prioritized Fictitious Self-Play}
In the experiments, we will evaluate the proposed method in a two-player boxing game. We are curious about how well an upper-level policy can reach by reusing the proposed NCP in a real-life sports game. To fulfill this, we employ prioritized fictitious self-play (PFSP) which has been demonstrated as a very effective multi-agent RL method in producing strong game AIs~\cite{vinyals2019grandmaster,han2020tstarbot}.

In PFSP, we launch parallel games, in each of which the current training agent and a historically dumped version of its policy parameters are chosen to perform a boxing game.
For every certain training iteration (around an hour), a copy of the agent's policy parameters is dumped and added to a candidate set of historical policies. 
At the beginning of a new game, the probability of sampling a historical policy $i$ from the candidate set $\mathcal{O}$ is given by:
\begin{equation} 
p_i=\frac{g(P_i)}{\sum_{o \in \mathcal{O}} g(P_o)},
\end{equation}
where $g:[0,1] \rightarrow[0, \infty)$ is a weighting function and $P_o\in[0,1]$ denotes the probability that the agent can defeat the policy $o$. Similar to Section \ref{sec:prioritized_sampling}, we also choose $g(x)=(1-x)^{\alpha_2}$.
Under this setup, the current player tends to choose the most challenging opponents in priority from the candidate set.




\section{RESULTS}
We evaluate the proposed framework on two benchmark motion datasets \cite{peng2022ase, won2021control}. We aim to verify the following statements:
1) our framework can learn a compact and informative latent representation that allows the characters to perform diverse and sophisticated movements, while maintaining realistic motion from the unstructured motion clips; 2) the shifted prior can generate various random trajectories covering the majority of the motions demonstrated in motion clips; 3) the upper-level policy can effectively reuse the pre-trained representation and prior distribution to solve new downstream tasks, including both single-agent task and two-player zero-sum game.

\subsection{Experimental Setup}
The simulation environments are implemented using the Isaac Gym simulator~\cite{makoviychuk2021isaac} with a simulation frequency of 120Hz, where the control policies execute at a frequency of 30Hz. 
To demonstrate the generality of our framework, we evaluated it on two different humanoid characters: one equipped with a sword and a shield with 37 degrees-of-freedom, similar to the character used in~\cite{peng2022ase}. Another character is equipped with boxing gloves with 34 degrees-of-freedom, similar to the character used in~\cite{won2021control}. 
The imitation policy for the sword \& shield character is trained on a total of approximately 26 minutes of motion data, consisting of 82 motion clips from \cite{peng2022ase} for moving and stunt motions, and 76 locomotion motion clips. For the boxer character, 45 boxing and locomotion motion clips from  \emph{\url{http://mocap.cs.cmu.edu}} and \emph{\url{http://mocap.cs.sfu.ca}} with their mirrored data are used, approximately 30 minutes in total.

All experiments are implemented on NVIDIA TESLA P40 GPUs under TLeague~\cite{sun2020tleague}, an efficient distributed multi-agent RL infrastructure. 
We use proximal policy optimization (PPO) \cite{schulman2017proximal} as the RL algorithm. For detailed hyperparameters, please refer to Section~\ref{appendix:hyperparameters} in Appendix. 
Both the imitation tasks for the sword \& shield character and boxer character take around 2 days of training. The prior training requires 12 hours. 
The strike task can be accomplished within 4 hours, while the boxing task demands 4 days to converge.

\subsection{Setup and Results for Imitation Policy}
\label{subsec:Low-Level}
\subsubsection{States}
In our framework, the state $s_t$ characterizes the proprioceptive observation of the character, which is defined as
\begin{equation}\label{eq:prop}
    s_t=\left\{q_j,\dot{q}_j, p_i, q_i, v_i, \omega_i, h\right\},
\end{equation}
where $q_j$ and $\dot{q}_j$ are the joint position and joint velocity of the $j$-th joint, respectively; $p_i$, $q_i$, $v_i$ and $\omega_i$ are the relative position, orientation, the linear and angular velocity of the $i$-th body expressed in the local coordinate frame of the root; $h$ represents the height of the root relative to the ground. The orientation is represented using a 6D vector~\cite{zhou2019continuity}.
In addition to the current proprioceptive observation, the imitation policy additionally takes input the future information vector $f_t$, which represents the future states of the character in motion clips. 
In our experiment, we define $f_t = \left\{\delta\left(s_t, s_{t+1}^*\right), \delta\left(s_t, s_{t+2}^*\right) \right\}$ to allow the imitation policy to receive the information of the subsequent two frames in motion clips. 
$\delta\left(s_t, s_{t}^*\right)$ is a function that measures some key information in the state $s_t$ and a target state $s_{t}^*$. It is defined as
\begin{equation}
    \delta\left(s_t, s_{t}^*\right)=\left\{q_j^*,\dot{q}_j^*,p_r^*,q_r^*,v_r^*,\omega_r^*,p_k^*\right\},
\end{equation}
where $q_j^*$ and $\dot{q}_j^*$ are target joint position and joint velocity of the $j$-th joint expressed in datasets; $p_r^*$, $q_r^*$, $v_r^*$, and $\omega_r^*$ are the relative position, orientation, linear and angular velocity of the target state expressed in the simulated character root coordinate frame; $p_k^*$ represents the position of certain key body parts of the target state represented in the target root coordinate frame. The key body parts for the boxer character are the hands and feet, while the other character utilizes both their hands and feet as well as a sword and shield.

The character with sword \& shield has a state $s_t$ with 362 dimensions and a future information vector $f_t$ with 284 dimensions to include information about the sword and shield, while the boxer character receives a state $s_t$ with 323 dimensions and a future information vector $f_t$ with 254 dimensions. 

\subsubsection{Actions}
The characters are immediately controlled by the output of the PD controllers, i.e., the torques. 
The control policies output target positions of the joints, which are fed into the PD controllers.
To alleviate the burden of exploration, the action $a_t$ serves as a residual target position. 
The target joint position $\hat{q}_t$ is calculated by adding the action vector $a_t$ to the current joint position $q_t$, i.e., $\hat{q}_t=q_t+a_t$. The character with sword \& shield has an action space with 31 dimensions, while the boxer character has an action space with 28 dimensions.

\subsubsection{Latent}
The codebook $e \in R^{K \times D}$ serves as the latent representation, where $D$ is the dimensionality of latent embedding vector $e_i$, and $K$ is the number of latent embedding vector. 
Throughout our experiments, 
we set $D=64$. 
For the character with sword \& shield, we set $K=512$, and we set $K=256$ for the boxer character.

\subsubsection{Network Architecture}
For the implementation of RL, we use the actor-critic infrastructure with PPO algorithm. Therefore, we need to train a policy (containing the encoder and decoder) for the actor and a value function for the critic.
The value function, encoder and decoder are all parameterized as deep neural networks. 
The value function $V(s_t, f_t)$ is modeled with a fully connected network with 3 hidden layers of $[1024, 512, 256]$ units. For the policy network, the encoder takes in state $s_t$ and $f_t$, and maps them to a latent variable $z_t^e$ using a fully connected network with 3 hidden layers of $[1024, 1024, 512]$ units. The latent $z_t^e$ is quantized to $z_t^q$, and the decoder takes in state $s_t$ and $z_t^q$ and maps them to a Gaussian distribution with a mean $\pi^{d}(a_t|s_t,z_t^q)$ and a trainable diagonal covariance matrix $\Sigma_\pi$. The decoder is also modeled with a fully connected network with 3 hidden layers of $[1024, 1024, 512]$ units. 

\subsubsection{Reward}
For the imitation policy, an objective similar to~\cite{peng2018deepmimic} is employed in our framework to track motion clips. The objective is
\begin{equation}\label{eq:tracking_objective}
    r_t=w^{jp} r_t^{jp}+w^{jv} r_t^{jv}+w^k r_t^k+w^r r_t^r+w^v r_t^v 
\end{equation}
where $r_t^{jp}$, $ r_t^{jv}$, $r_t^k$, $r_t^r$, and $r_t^v$ correspond to the tracking reward of joint angles, joint velocities, key body, root, and root velocity, respectively. The weight of each reward is denoted as $w$ with the same superscript.
See Section~\ref{appendix:motion_tracking_objective} in Appendix for detailed information on reward settings.

\begin{figure}[t]
  \centering
  \includegraphics[width=\linewidth]{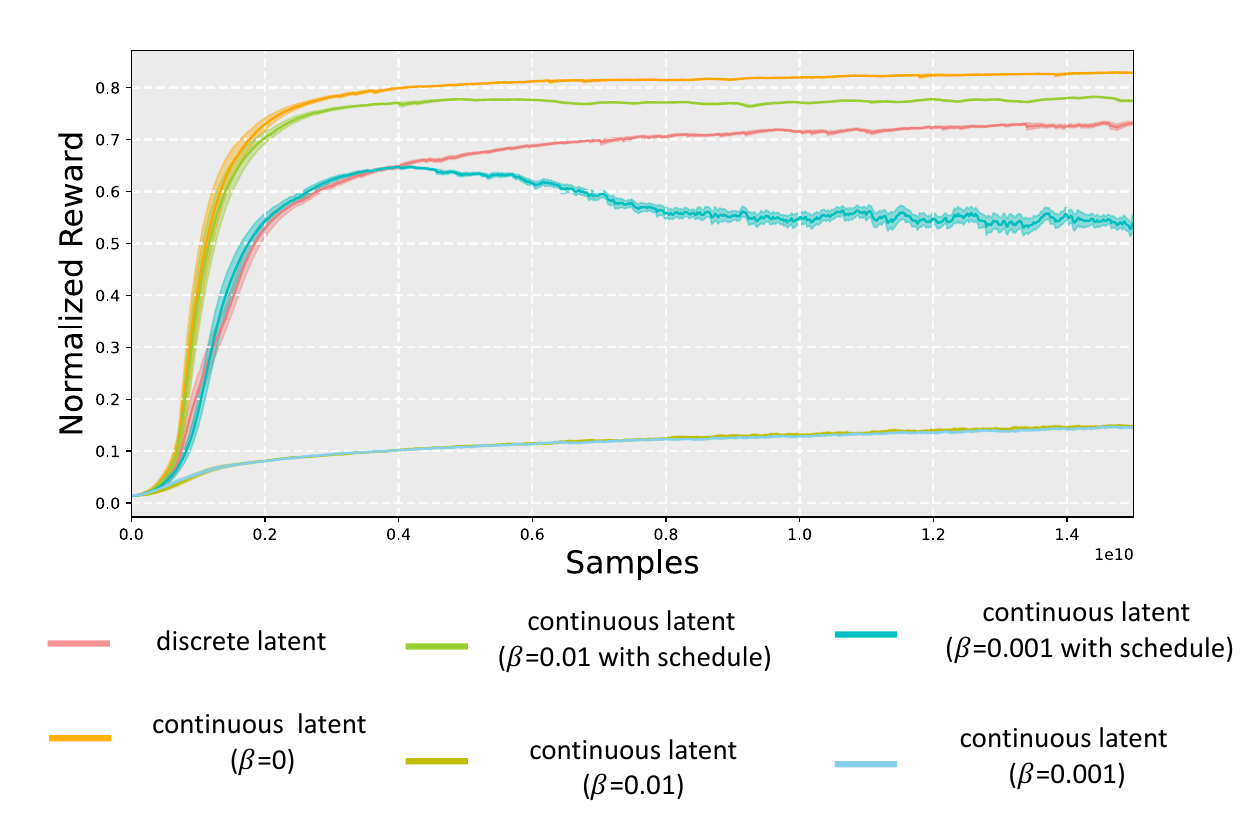}
  \caption{Learning curves of various $\beta$-VAE variants and ours (denoted as discrete latent).}
  \label{fig:tracking_curve}
\end{figure}

\begin{figure}[t]
  \centering
  \includegraphics[width=\linewidth]{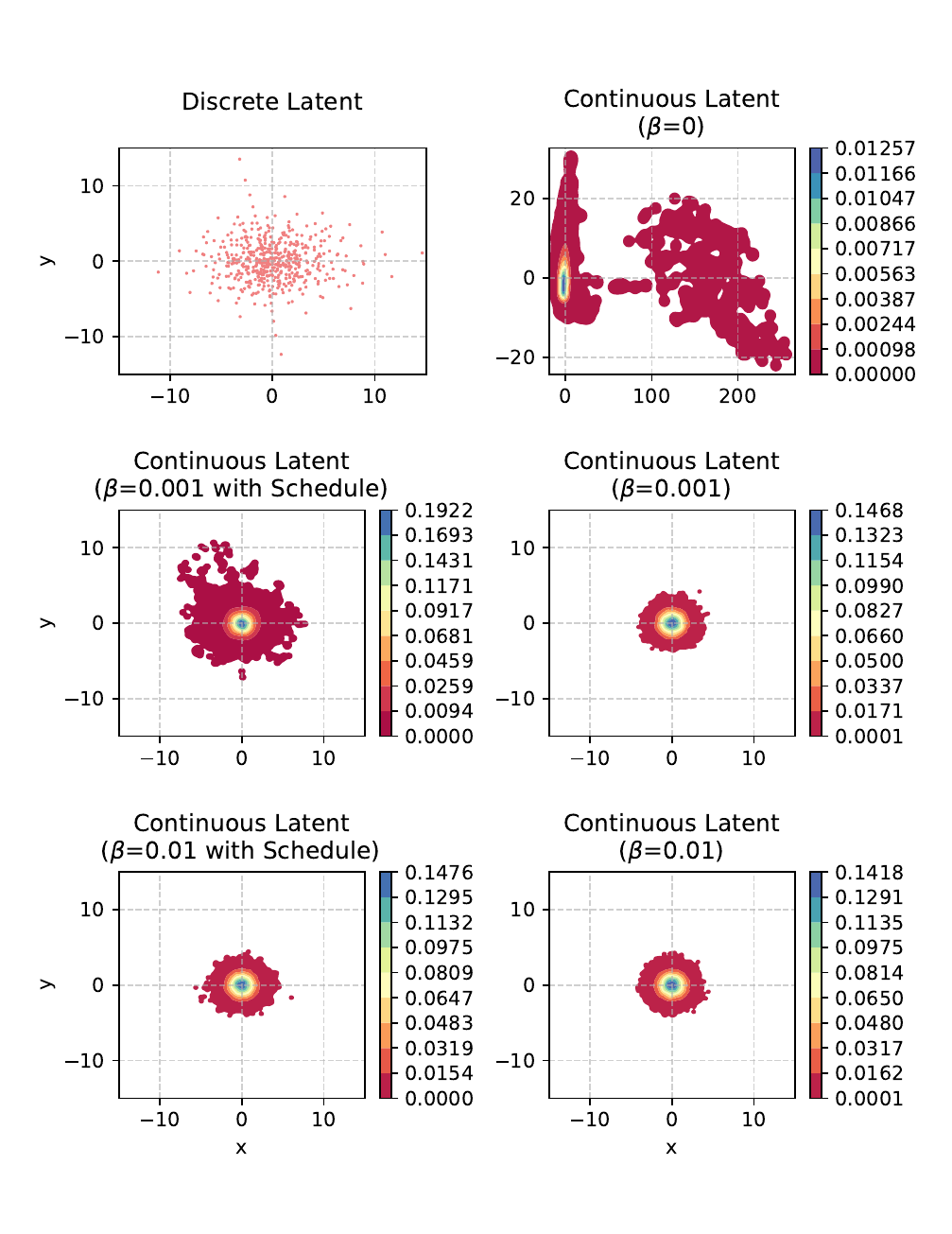}
  \caption{Visualization of discrete and continuous latent distributions from various methods using Principal Component Analysis (PCA). For the continuous latent distribution, Kernel Distribution Estimation (KDE) is employed to estimate the distribution.}.
  \label{fig:latent}
\end{figure}

\subsubsection{Evaluation}
We train each humanoid character with an imitation policy.
The results show that each character is able to perform sufficiently diverse and realistic movements, including locomotion, boxing, and sword \& shield stunts. 
To quantitatively analyze the effectiveness of the imitation policy, we use the sword \& shield dataset and conduct comprehensive comparative experiments with VAE-based- and GAN-based methods below.
\begin{figure*}[t]
  \centering
  \includegraphics[width=\linewidth]{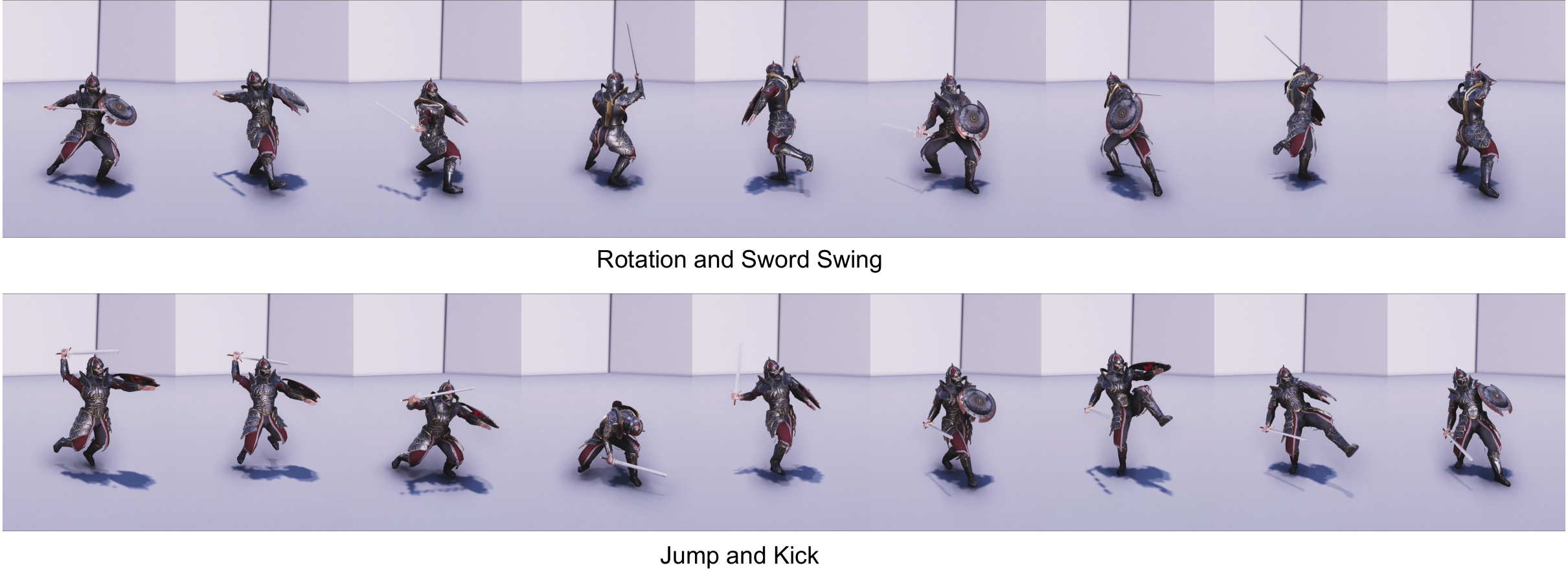}
  \caption{The tracking performance of the imitation policy of NCP for the character with sword \& shield.}
  \Description{test}
  \label{fig:TOG_tracking}
\end{figure*}

\begin{figure*}[t]
  \centering
  \includegraphics[width=\linewidth]{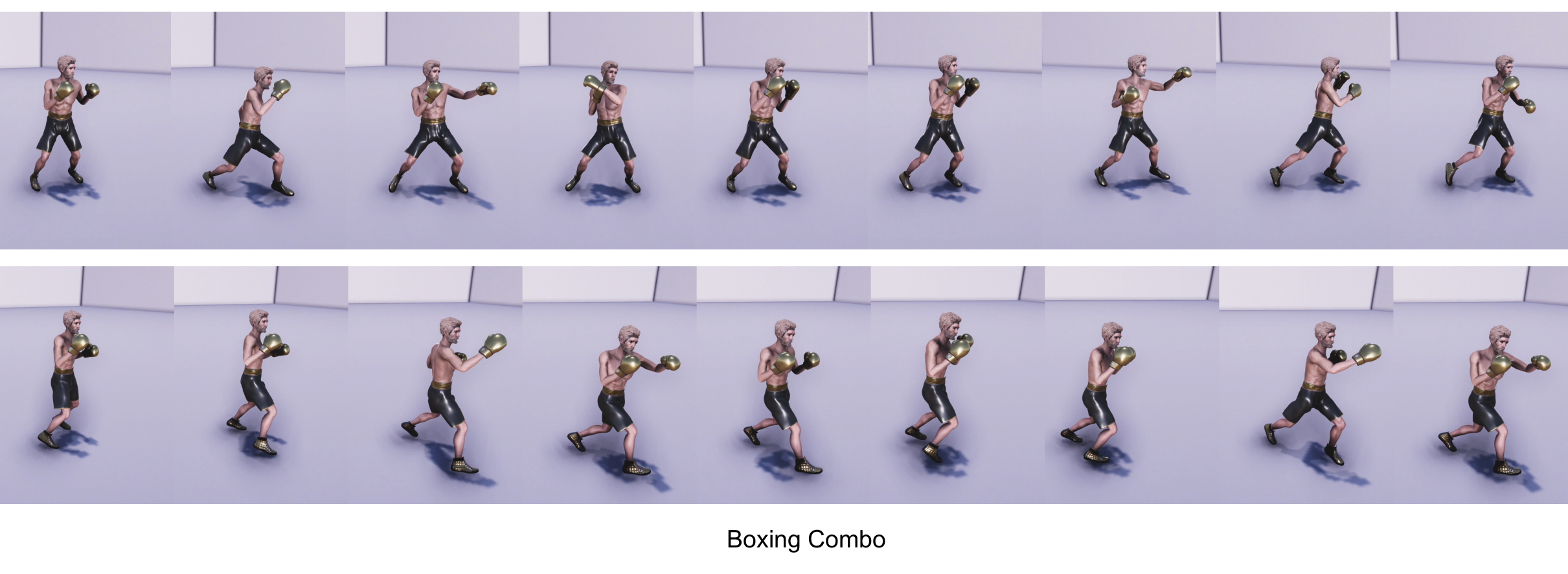}
  \caption{The tracking performance of the imitation policy of NCP for the boxer character.}
  \Description{test}
  \label{fig:boxing_tracking}
\end{figure*}

\noindent
\textbf{Comparison with VAE-based methods.} We compare with a number of $\beta$-VAE variants with different settings on $\beta$ in Eq.~\eqref{vae-objective}. 
Specifically, we test $\beta=0.01$ and $\beta=0.001$ with and without the annealing schedule trick, 
resulting in 4 $\beta$-VAE candidates. Despite those, we additionally consider an extreme case with $\beta=0$.
For a fair comparison, for these $\beta$-VAE variants we use similar neural network architecture as ours and the dimension of their latent variable $z$ is set to $64$ as well.
The reward curves comparing the tracking performance of all candidates are shown in Fig~\ref{fig:tracking_curve}. 
At the same time, we would like to understand the expressive power of the latent representations for different methods, since VAE-based methods including ours try to seek for a balance between the reconstruction quality and the compatibility of the latent distribution with a prior.
To understand the latent space, we plot the 64 dimensional latent variables for all the compared methods on a 2D space using Principal Component Analysis (PCA). The visualization of the 2-dimensional latent is shown in Fig.~\ref{fig:latent}.

From Figs.~\ref{fig:tracking_curve} and~\ref{fig:latent}, as expected, the $\beta=0$ variant shows the best tracking performance, since the latent space is not regularized; however, its latent distribution is broad and apparently not reusable for action generation. 
For $\beta=0.01$ and $\beta=0.001$, both variants cause posterior collapse, resulting in latents that are close to the standard Gaussian distribution while they fail to track the motion clips. 
On the other hand, the annealing schedule~\cite{bowman2015generating} trick is verified effective in preventing posterior collapse, using which $\beta$ linearly increases from 0 to 0.01 or 0.001 during training. The according two variants with annealing schedule achieve a relatively compact latent space and maintain a relatively good tracking performance.

The empirical results on the continuous latent space demonstrate again that increasing $\beta$ could produce a compact latent space, but ultimately lead to deteriorated tracking performance, while decreasing $\beta$ can enhance tracking performance but lead to a widely distributed latent distribution, making it challenging for upper-level policy usage. 
Therefore, it is not trivial for VAE-based methods to reach a perfect balance. 

Now, we discuss the performance of our proposed method. 
According to Fig.~\ref{fig:tracking_curve}, the tracking performance of the proposed method is slightly lower than the $\beta=0$ VAE variant and $\beta=0.001$ with annealing schedule variant. As demonstrated in the videos, such tracking performance is sufficient to produce high-quality movements.
For the learned latent space, it is obvious in Fig.~\ref{fig:latent} that our method produces a discrete latent space with the scattered codes indicating the means of multiple Gaussians. 
The codes span over the 2D space and provide a broad yet compact distribution, since the discrete space is quite small, with only $256$ / $512$ codes for the boxer / sword \& shield character.
Unlike VAE, the discrete latent space considerably facilitates upper-level usage. For the case of upper-level RL, it alleviates the exploration burden of the upper-level policy, compared to exploring in a continuous Gaussian space. 

\noindent
\textbf{Comparison with GAN-based methods.} Recent developments in GAN-based methods have shown impressive results in generating high-quality skills for physics-based characters. To showcase the effectiveness of our framework, we compare our results with the state-of-the-art approach ASE~\cite{peng2022ase} and CALM~\cite{tessler2023calm} using the same motion datasets. 
Both ASE~\cite{peng2022ase} and CALM~\cite{tessler2023calm} have released the trained model parameters, and hence we direct evaluate on these models without any bias.
To achieve a fair comparison, we first generate approximately one million data in $\mathcal{F}$ by randomly sampling from the latent space of each method in ASE, CALM, and our NCP, separately.
Then, we define a general reconstruction score $s(f_i^*)$ for frame $f_i^*$ in motion clips as
\begin{equation}
s(f_i^*) = \underset{f \in \mathcal{F}}{\max } \quad 0.5 \cdot r^{jp}(f_i^*, f) + 0.5 \cdot r^v(f_i^*, f),
\end{equation}
where $r^{jp}(f_i^*, f)$ and $r^v(f_i^*, f)$ are defined identically as in Eq.~\eqref{eq:tracking_objective}. $r^{jp}$ and $r^v$ measure the similarity of joint position and root velocity between frame $f_i^*$ stored in motion clips and frame $f$ sampled from the decoder of each method.
Note that none of the ASE, CALM and ours directly optimize the above score, and the score is general to evaluate the performance of generated motions from various control policies.


\begin{figure}[t]
  \centering
  \includegraphics[width=\linewidth]{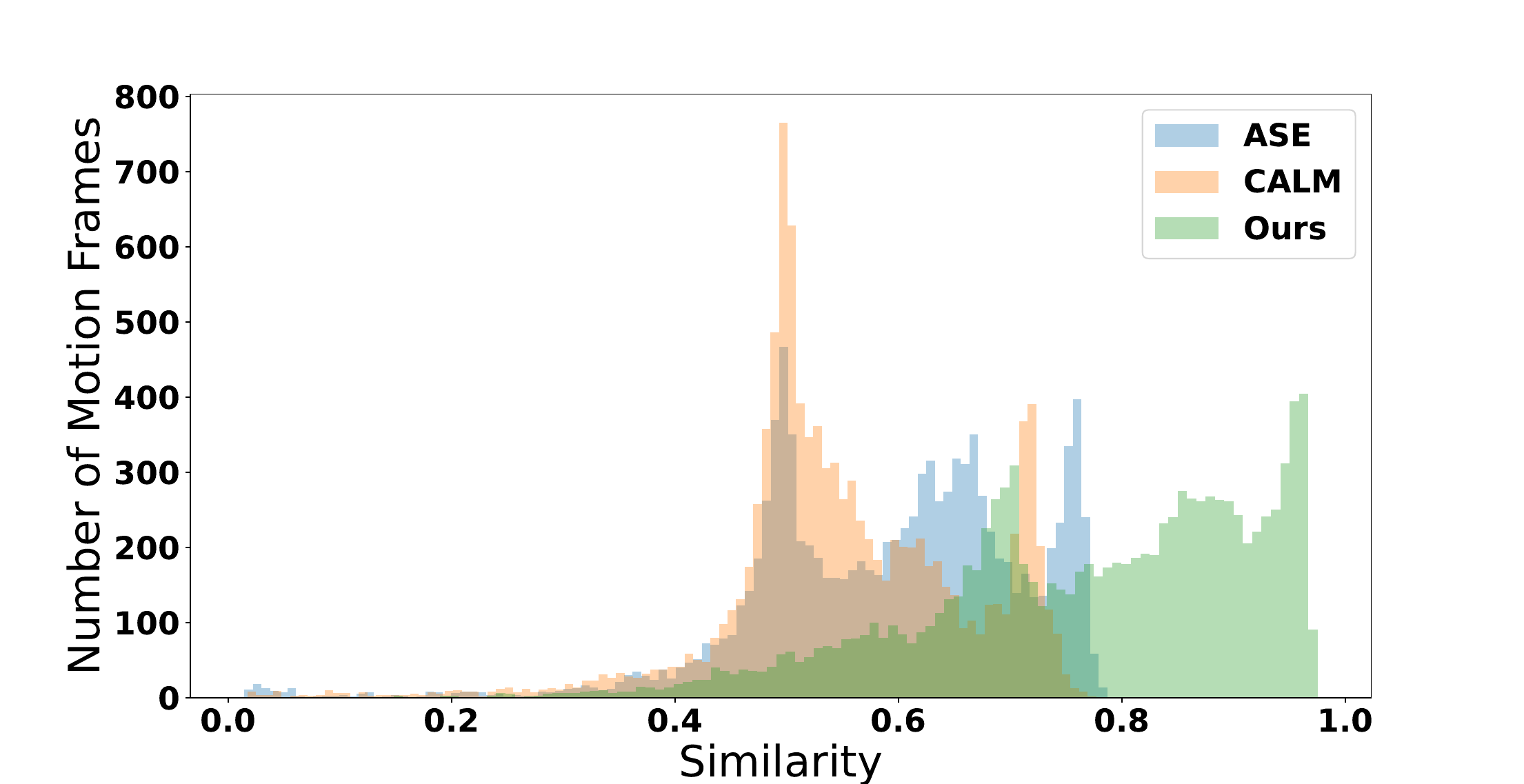}
  \caption{Comparison of reconstruction similarity scores of ASE, CALM, and our method.}.
  \label{fig:heat}
\end{figure}

The above reconstruction scores for different methods are evaluated in Fig.~\ref{fig:heat}, where
the $y$-axis indicates the number of motion frames and the $x$-axis indicates the computed reconstruction similarity score.
Intuitively, more motion frames falling in the interval with higher scores indicate better reconstruction performance and motion coverage. 
That is, the histogram area centered at the right-hand side indicates better performance.
As observed in the figure,
CALM exhibits better diversity and tracking performance compared to ASE, and this is consistent with the conclusion in~\cite{tessler2023calm};
while our model produces significantly more naturalistic and diverse behaviors compared to both ASE and CALM. This conclusion can also be obviously observed from the videos in the supplementary materials.
Furthermore, we count the number of frames that obtain a reconstruction score above 0.5 and compute the ratio over the total number of training frames.
Our approach achieved a rate of 94.6\%, whereas ASE achieved 77.1\% and CALM achieved 69.6\%. 

Finally, we screenshot a few frames of the simulated characters when tracking the motion clips in Figs.~\ref{fig:TOG_tracking} and~\ref{fig:boxing_tracking}. As shown in these images, both the characters can perfectly reproduce some complex movements in the motion clips, such as rotation, sword swing, jumping and kicking movements for the sword \& shield character, and combination of punches for the boxer character.

\subsection{Setup and Results for Prior}

\subsubsection{Observation and Action}
In our framework, the prior only takes proprioceptive observation $s_t$ of the character as input, as defined in Eq.~\eqref{eq:prop}. The action space of the prior is the categorical distribution over the codes. In the case of the sword \& shield character, the action is represented as a 512-dimensional vector, i.e., the code, while for the boxer character, it is a 256-dimensional vector.

\subsubsection{Network Architecture}
During the training of prior, both the value function and policy function are modeled using fully connected neural networks. The value function consists of three hidden layers with 1024, 512, and 256 units, respectively. Similarly, the policy function is modeled using three hidden layers, consisting of 1024 units in each layer.

\begin{figure}[h]
  \centering
  \includegraphics[width=\linewidth]{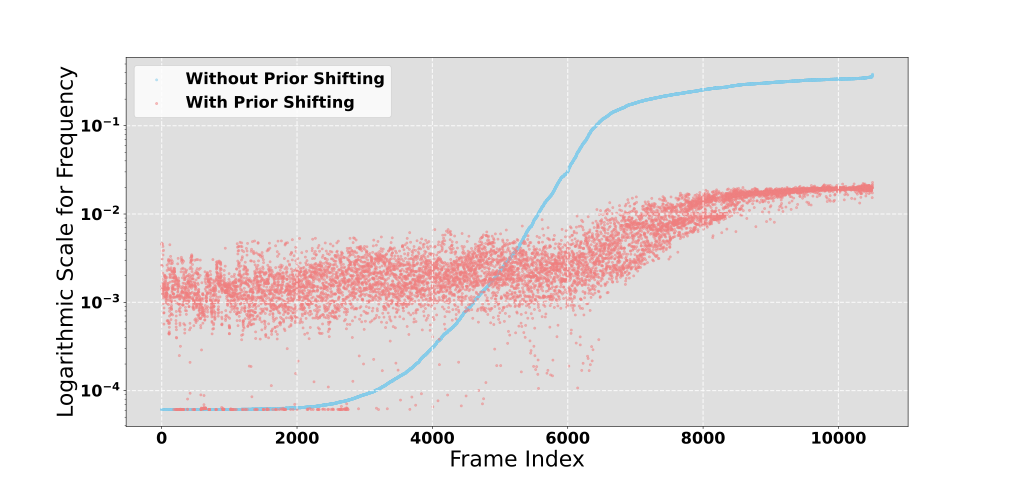}
  \caption{Comparison of frame visit frequency between without prior shifting and with prior shifting for the character equipped with sword \& shield.}.
  \label{fig:ss_prior}
\end{figure}
\begin{figure}[h]
  \centering
  \includegraphics[width=\linewidth]{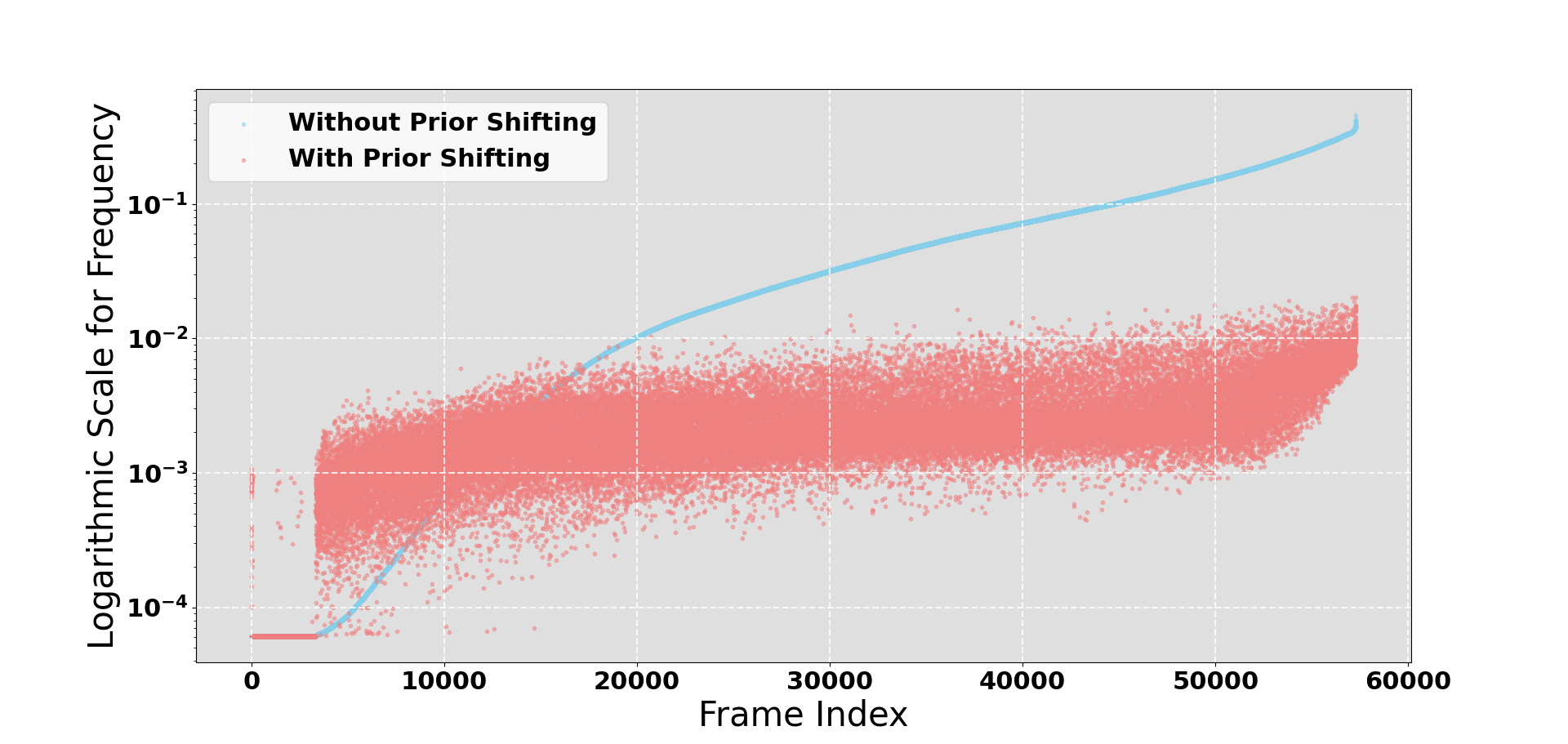}
  \caption{Comparison of frame visit frequency between without prior shifting and with prior shifting for the boxer character.}.
  \label{fig:boxing_prior}
\end{figure}

\subsubsection{Evaluation}
In this section, we evaluate the effectiveness of the proposed prior shifting technique. Since prior shifting is introduced to produce more diverse behaviors, we perform statistics on the generated movements by randomly sampling codes from the priors with and without prior shifting. 

Figs.~\ref{fig:ss_prior} and~\ref{fig:boxing_prior} show the comparison of frame visitation frequencies (with log scale) of using and not using the prior shifting for the sword \& shield character and boxer character, respectively. The frames along the $x$-axis are sorted by their visitation frequencies before prior shifting, and hence the blue points show a smooth pattern along the $x$-axis.
As we can observe, using prior shifting, the shifted prior distribution is tuned to match a nearly even distribution over all the motion frames. This provides sufficient confidence for reusing this prior in downstream tasks.

\subsection{Setup and Results for Upper-Level Policy}

\subsubsection{Observation and Action}
The upper-level policy $\pi_{\vartheta}^h(\cdot| s_t,g_t)$ takes the proprioceptive observation $s_t$ defined in Eq.~\eqref{eq:prop} and a task-specific goal $g_t$ as inputs. 
The goal $g_t$ for the strike task is defined similarly to~\cite{peng2022ase}, using a 15-dimensional vector that represents the relative position, orientation, linear velocity, and angular velocity of the target. 
While in the boxing task, the goal $g_t$ represents the opponent's state by a 289-dimensional vector that includes the body position, rotation, root linear velocity, and angular velocity of the opponent, expressed in the root coordinate system of the agent.
The action space of the upper-level policy is consistent with the prior network which the policy selects a code from the categorical distribution to drive the decoder to generate movements. The action of the sword \& shield character is represented by a 512-dimensional vector, while that of the boxer character is a 256-dimensional vector.

\subsubsection{Network Architecture}
Both the value function and policy function in the upper-level policy are modeled using fully connected neural networks. The value function is composed of three hidden layers with 1024, 512, and 256 units, and the policy function is composed of three hidden layers with 1024, 512, and 512 units, respectively.
\begin{figure}[h]
  \centering
  \includegraphics[width=\linewidth]{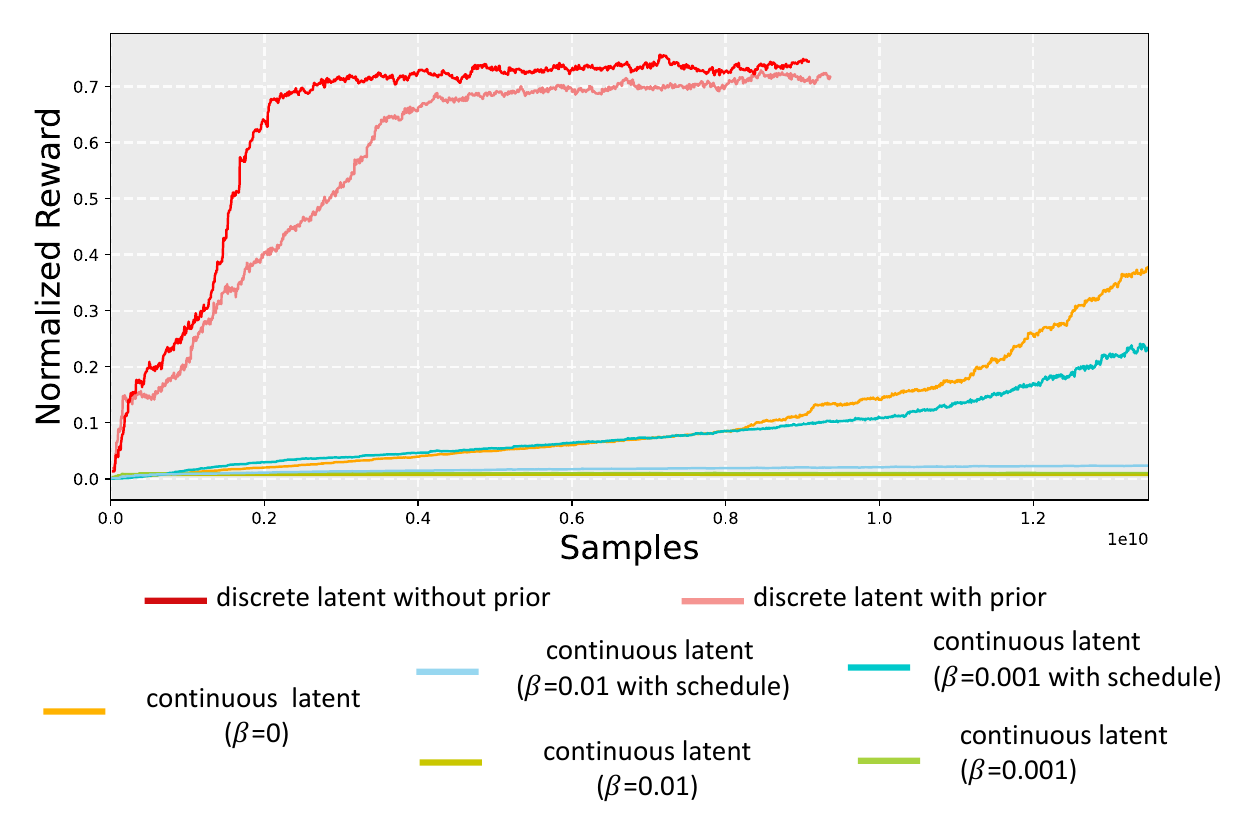}
  \caption{Upper-Level policy learning curve using different decoders.}.
  \label{fig:high-level_learning_curve}
\end{figure}

\begin{figure}[h]
  \centering
  \includegraphics[width=\linewidth]{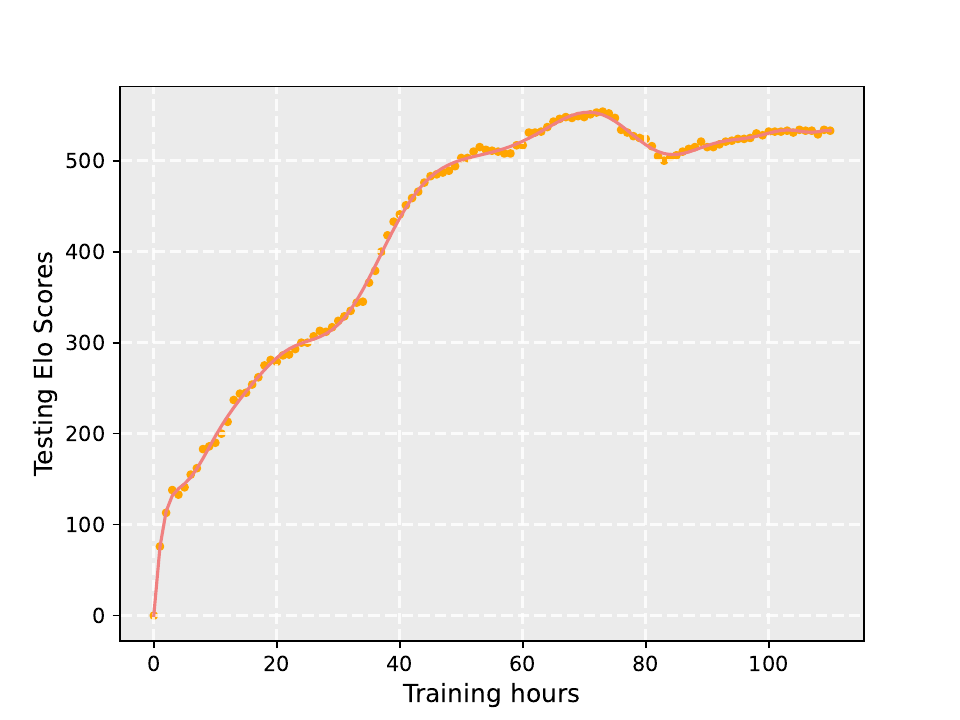}
  \caption{The testing Elo scores of generated players during the PFSP training. 
  Each point represents the score of a trained player along the training time.}
  \Description{test}
  \label{fig:elo_score}
\end{figure}

\begin{figure*}[t]
  \centering
  \includegraphics[width=\linewidth]{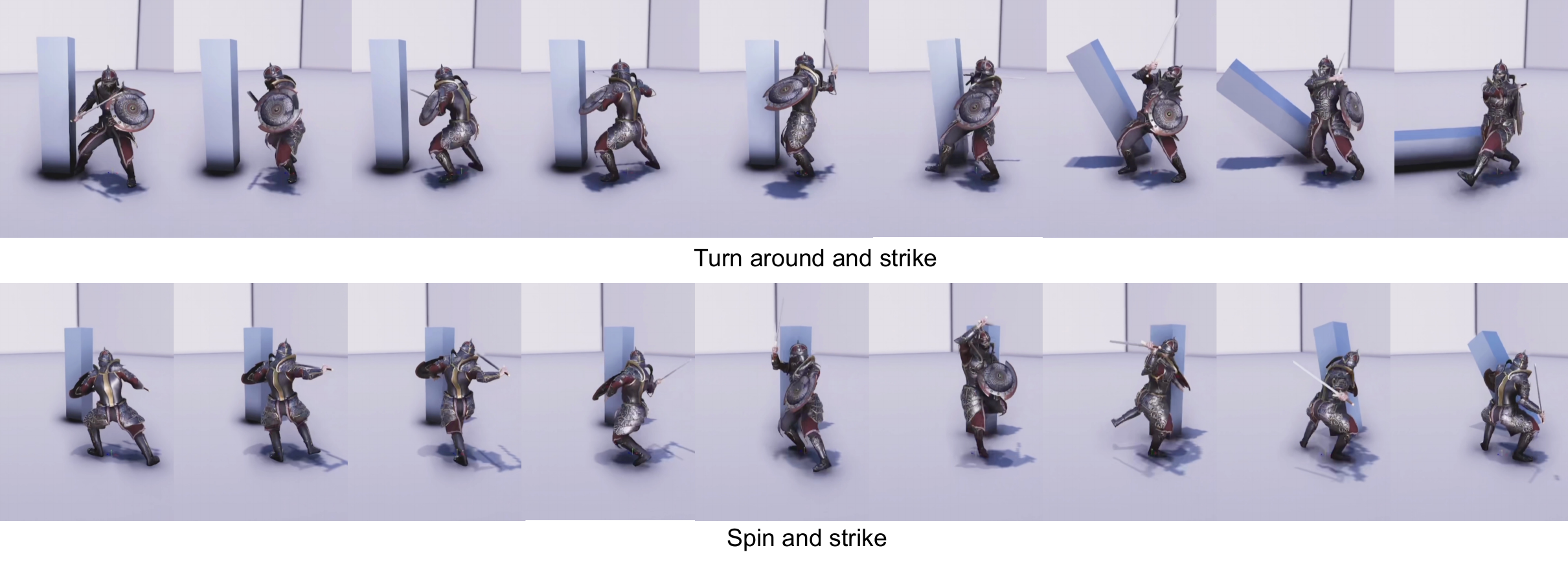}
  \caption{The performance of our method in the strike task.}
  \Description{test}
  \label{fig:TOG_strike}
\end{figure*}

\begin{figure*}[t]
  \centering
  \includegraphics[width=\linewidth]{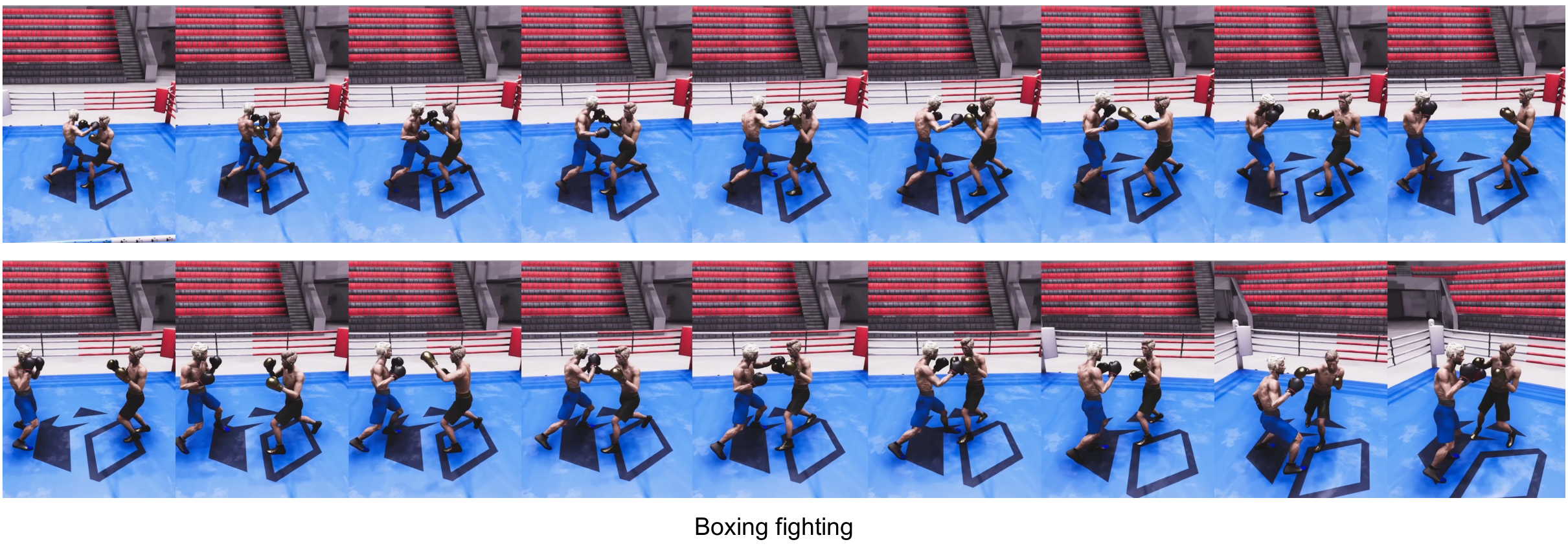}
  \caption{The performance of our method in the two-player boxing game.}
  \Description{test}
  \label{fig:boxing_fight}
\end{figure*}

\subsubsection{Reward}
The reward function for the strike task is similar to that for ~\cite{peng2022ase}:
\begin{equation*}
r=w_{\text {strike }} r_{\text {strike }}+w_{\text {facing }} r_{\text {facing }}+w_{\text {vel }} r_{\text {vel }},
\end{equation*}
and the strike reward $r_{\text {strike }}$ is defined as
\begin{equation*}
r_{\text {strike }} = 1 - u^{up} \cdot  u_t^*,
\end{equation*}
where $u^{up}$ and $u_t^*$ represent the up vector of the global coordinate and the target object, respectively. The facing reward $r_{\text {facing}}$ is defined as 
\begin{equation*}
\begin{aligned}
& r_{\text {facing }}=\exp (-5\cdot|1-\mathbf{d} \cdot \hat{\mathbf{d}}|), \\
& {\mathbf{d}}=\left(\mathbf{p}_{\mathrm{t}}-\mathbf{p}\right) /\left\|{\mathbf{p}}_{\mathrm{t}}-{\mathbf{p}}\right\|,
\end{aligned}
\end{equation*}
where $\hat{\mathbf{d}}$ is the facing direction of the character, $\mathbf{p}$ and $\mathbf{p}_{\mathrm{t}}$ are the global position of the character and target separately.
The velocity reward $r_{\text {vel}}$ is defined as
\begin{equation*}
r_{\mathrm{vel}}=\exp \left(-4\cdot\left|\mathbf{v^*}-\mathbf{d} \cdot \mathbf{v}\right|\right),
\end{equation*}
where $\mathbf{v}$ and $\mathbf{v^*}$ are the global velocity and the goal velocity of the player, respectively. The weights for the strike, facing, and velocity rewards are represented by $w_{\text {strike }}$, $w_{\text {facing }}$, and $w_{\text {vel }}$, respectively. In our experiments, these weights are set to 0.6, 0.2, and 0.2, respectively.

The reward function for the boxing task is designed similarly to~\cite{won2021control} as
\begin{equation*}
r=r_{\text {damage }}+w_{\text {facing }} r_{\text {facing }}+w_{\text {vel }} r_{\text {vel }}+w_{\text {fall }} r_{\text {fall }},
\end{equation*}
and the damage reward $r_{\text {damage }}$ is defined as 
\begin{equation*}
r_{\text {damage }} = \left\|f_{pl->op}\right\|-\left\|f_{op->pl}\right\|,
\end{equation*}
where $\left\|f_{pl->op}\right\|$ is the contact force between the hands of agent and the opponent to measure the damage of punches against the opponent. 
Different from~\cite{won2021control}, the contact force in our task is clipped between 200N and 1200N, which is a natural range for human boxing players. The facing reward $r_{\text {facing}}$ and velocity reward $r_{\text {vel}}$ are defined similarly to the strike task, with the difference being that the target position is now the opponent's root position.
 Additionally, a sparse binary reward $r_{\text {fall}}$ is designed for the agent when the opponent falls down. In our experiment, the weights for $w_{\text {facing }}$, $w_{\text {vel}}$, and $w_{\text {fall}}$ are set to 0.6, 0.4, and 100, respectively.

\subsubsection{Evaluation}
In the last part of the experiments, we evaluate the performance of our method for solving downstream tasks. As mentioned previously, we consider two tasks that are strike task for the character with sword \& shield and two-player boxing game.
To support the effectiveness of the pre-trained discrete latent representations and prior networks in promoting the efficiency of upper-level policy training, we propose a few ablation studies here to test the decoders trained from various $\beta$-VAE methods and ours.

The training curves of the upper-level policies using different decoders are reported in Fig.~\ref{fig:high-level_learning_curve} for the strike task. 
As expected, the results show that using the discrete latent can significantly facilitate and speed up the RL training process. 
Another conclusion in the curves is that without using the trained prior distribution as a constraint in the KL-regularized term, the policy achieves a slightly higher reward than the KL-regularized version with the prior. 
This is reasonable because without any constraint, the upper-level policy is free to explore the discrete latent space with a single objective that is maximizing the task reward. However, this might cause the policy fails to perform diverse strategies that can elaborate all movements in the motion clips.
This is verified in our videos, where the trained policy with prior regularization demonstrates much more diverse behaviors compared with the one without prior regularization.
Fig.~\ref{fig:TOG_strike} shows some screenshots of the strike task, where the character could carry out a diverse set of behaviors to strike the square pillar, compared to the performance of the ASE method~\cite{peng2022ase} in the same task.

In the two-player boxing game, we additionally consider a multi-agent RL problem and train the upper-level policy using the PFSP algorithm. 
The training lasts for around 4.5 days. During this time, we dump the network parameters every hour, and a total number of 110 models are stored.
To witness how the training progress proceeds, 
we evaluate the 110 models with Round Robin tournament and perform 100 matches for each pair of players to compute the final payoff matrix. The Elo scores~\cite{elo} (a common metric for measuring the overall performance of a player in Round Robin tournament)
are plotted in Fig.~\ref{fig:elo_score} and demonstrate that the trained boxer can consistently improve and overcome past models.


Finally, we screenshot some impressive frames of the simulated characters in the two-player boxing game in Fig.~\ref{fig:boxing_fight}. Surprisingly, we observe that the character emerges life-like strategies similar to professional human boxers in boxing matches, such as defense and dodge. More details can be seen in the video.


\section{DISCUSSION}
This paper introduces a novel learning framework that enhances the quality and diversity of physics-based character control beyond current state-of-the-art methods. Our approach employs reinforcement learning (RL) to first track and replicate realistic movements from unstructured motion clip datasets using the discrete information bottleneck.
This structure extracts the most pertinent information from motion clips into a compact yet informative discrete latent space. 
By sampling latent variables from a trained prior distribution, we can generate high-quality lifelike behaviors. 
However, this prior distribution relies on the original data distribution and could be affected by motion imbalance in the dataset. 
To address this issue, we further propose a technique called prior shifting to adjust the trained prior using curiosity-driven RL. 
The shifted distribution offers a nearly even distribution of the generated movements over the motion clips in the dataset. 
This enables its easy usage in upper-level policy, and even a random upper-level policy can produce ample behavioral diversity.


In future work, we aim to investigate the potential of adapting this framework to environmental changes, such as accommodating additional objects and characters. This could enable us to expand the learned policy to adapt to previously unseen environments.
Similar to all data-driven methods, 
the performance of our system is limited by the size of the training dataset.
Besides, enhancing our ability to learn more complex behaviors as well as extending the framework to larger datasets, promise to be a fascinating area for future exploration.
%
%
%
%

\bibliographystyle{ACM-Reference-Format}
\bibliography{bibliography}

\clearpage
\appendix
\section{hyperparameters}
\label{appendix:hyperparameters}

Table~\ref{tab:hyperparameters_imitatoin} lists the hyperparameter settings used for the imitation policy, while Table~\ref{tab:hyperparameters_prior} provides the corresponding hyperparameters for prior shifting. Finally, Table~\ref{tab:hyperparameters_upper} contains the hyperparameters for the upper-level policy.

\begin{table}[H]
\centering
	\caption{Hyperparameters for training imitation policy.}
 \label{tab:hyperparameters_imitatoin}
\begin{tabular}{|l|p{40pt}|} 
\hline
\textbf{Hyperparameter}                                & \textbf{Value}      \\ 
\hline
Number of Code $K$                                     & 512 or 256      \\ 
\hline
Code Dimension $D$                                     & 64       \\ 
\hline
Commitment Penalty $\beta$                             & 0.25       \\ 
\hline
GAE($\lambda$)                                         & 0.95     \\ 
\hline
Discount Factor $\gamma$                               & 0.95     \\ 
\hline
Policy/Value Function Learning Rate                    & 0.00005  \\ 
\hline
Policy/Value Function Minibatch Size                   & 16384    \\ 
\hline
PPO Clip Threshold                                     & 0.1      \\ 
\hline
Prioritized Sampling Coefficient $\alpha_1$              & 3        \\
\hline
\end{tabular}
\end{table}

\begin{table}[H]
\centering
	\caption{Hyperparameters for prior shifting.}
 \label{tab:hyperparameters_prior}
\begin{tabular}{|l|p{40pt}|} 
\hline
\textbf{Hyperparameter}                                & \textbf{Value}      \\ 
\hline
Number of Counted States $T_k$                         & 4915200     \\ 
\hline
Reward Scale Factor $N_k$                              & $2 \times \log(T_k)$     \\ 
\hline
GAE($\lambda$)                                         & 0.95     \\ 
\hline
Discount Factor $\gamma$                               & 0.95     \\ 
\hline
Policy/Value Function Learning Rate                    & 0.00001  \\ 
\hline
Policy/Value Function Minibatch Size                   & 4096    \\ 
\hline
PPO Clip Threshold                                     & 0.1      \\ 
\hline
\end{tabular}
\end{table}

\begin{table}[H]
\centering
	\caption{Hyperparameters for training upper-level policy.}
  \label{tab:hyperparameters_upper}
\begin{tabular}{|l|p{40pt}|} 
\hline
\textbf{Hyperparameter}                                & \textbf{Value}      \\ 
\hline
KL Loss Weight $\alpha_{\mathrm{KL}}$                  & 0.05       \\ 
\hline
Entropy Loss Weight $\alpha_{\mathrm{H}}$             & 0.01       \\ 
\hline
GAE($\lambda$)                                         & 0.95     \\ 
\hline
Discount Factor $\gamma$                               & 0.95     \\ 
\hline
Policy/Value Function Learning Rate                    & 0.00005  \\ 
\hline
Policy/Value Function Minibatch Size                   & 8192    \\ 
\hline
PPO Clip Threshold                                     & 0.1      \\ 
\hline
PFSP Sampling Coefficient $\alpha_2$              & 2        \\
\hline
\end{tabular}
\end{table}

\section{MOTION TRACKING OBJECTIVE}
\label{appendix:motion_tracking_objective}
In Section \ref{subsec:Low-Level}, we propose that our tracking reward is a weighted sum of specific reward terms. In this section, we provide a description of those terms. The reward $r_t$ is defined as
\begin{equation}
  \nonumber
    r_t=w^{jp} r_t^{jp}+w^{jv} r_t^{jv}+w^k r_t^k+w^r r_t^r+w^v r_t^v,
\end{equation}
\begin{equation}
  \nonumber
    w^{jp}=0.3, w^{jv}=0.1, w^{k}=0.3, w^{r}=0.2, w^{v}=0.1,
\end{equation}

The joint angles reward $r_t^{jp}$ and joint velocities reward $r_t^{jv}$ are defined as:
\begin{equation}
  \nonumber
  r_t^{jp}=\exp \left[-2.0\left(\sum_j\left\|q_{t, j}^*-q_{t, j}\right\|\right)\right],
\end{equation}

\begin{equation}
  \nonumber
  r_t^{jv}=\exp \left[-0.1\left(\sum_j\left\|\dot{q}_{t, j}^*-\dot{q}_{t, j}\right\|\right)\right],
\end{equation}
where $q_{t, j}$ and $\dot{q}_{t, j}$ represents the joint angle and joint velocity of the j-th joint at time t, while $q_{t, j}^*$ and $\dot{q}_{t, j}^*$ are the corresponding target joint angle and joint velocity represented in motion clips.

The key body reward $r_t^{k}$ is defined to tracking the position of the key body of the agent, which is formulated as follows
\begin{equation}
  \nonumber
  r_t^{k}=\exp \left[-10.0\left(\sum_k\left\|p_{t, k}^*-p_{t, k}\right\|\right)\right]
\end{equation}
where $p_{t, k}$ represents the position of the k-th body relative to the root, and $p_{t,k}^*$ represents the position in motion clips.

For the root, two types of rewards are defined as
\begin{equation}
  \nonumber
  r_t^{r}=\exp \left[-20.0\left(\left\|p_{t, r}^*-p_{t, r}\right\|^2 + 0.5 * \theta^2 \right)\right],
\end{equation}
 
\begin{equation}
  \nonumber
  r_t^{v}=\exp \left[-2.0\left(\left\|v_{t, r}^*-v_{t, r}\right\|^2 + 0.1 * \left\|\omega_{t, r}^*-\omega_{t, r}\right\|^2 \right)\right],
\end{equation}
where $p_{t, r}$, $v_{t, r}$, and $\omega_{t, r}$ represent the position, linear velocity, and angular velocity of the root, respectively, $p_{t, r}^*$, $v_{t, r}^*$, and $\omega_{t, r}^*$ are the target stored in motion clips. $\theta$ denotes the angle in the axis-angle representation of $R^* R^{-1}$, while $R^*$ represent the root orientation in motion clips and $R$ is the root orientation of the agent.

\section{EXPLANATION of EQUATION~(8)}
\label{appendix:explanation}

Without loss of generality, $p(\cdot)$ stands for a probability function without other specifications.
The probability of the trajectory $\tau$ in Eq.~\eqref{kl_loss} is
\begin{equation*}
\pi_{\varphi}^p(\tau)=p\left(s_0\right) \prod_{t=0} \pi_{\varphi}^{p}\left(k| s_t\right) \pi_{\phi}^{d}\left(a_{t} | s_t, (e_k)_t\right) p\left(s_{t+1} | s_t, a_t\right)
\end{equation*} 
where the index $k$ is sampled from the categorical prior network $\pi_{\varphi}^p$ and $(e_k)_t$ is the corresponding code $e_k$ at time step $t$. 
Then, Eq.~\eqref{kl_loss} can be derived as
\begin{equation}
\label{eq:vqvae-plogp}
\begin{aligned}
&\mathbb{E}_{\tau \sim \pi_{\varphi}^{p}, \atop f_t\sim\mathcal{M}}\ 
\sum_{t=0}\sum_{z_t^e\sim p_{\theta}^{e}} 
p_{\theta}^e(z_t^e | s_t,f_t)
\log\frac{p_{\theta}^e(z_t^e | s_t,f_t)}{\pi_{\varphi}^{p}(\mathbb{I}(z_t^e) | s_t)}
\\
=\ 
&\mathbb{E}_{s,f,z,a\ \sim \atop 
\rho^{p_{\theta}^e,\pi_{\varphi}^{p},\pi_{\phi}^{d},\mathcal{M}}
}
\ 
-
\pi_{\varphi}^{p}(\mathbb{I}(z) | s)\log\pi_{\varphi}^{p}(\mathbb{I}(z) | s),
\end{aligned}
\end{equation}
where the equality holds by noting that $p_{\theta}^e(z_t^e | s_t,f_t)$ has a probability mass on the nearest code according to Eq.~\eqref{vq-vae-posterior}; $\rho^{p_{\theta}^e,\pi_{\varphi}^{p},\pi_{\phi}^{d},\mathcal{M}}$ is a visitation probability commonly used in RL theories. Eq.~\eqref{eq:vqvae-plogp} demonstrates that optimizing the distillation loss acts similarly with the prior fitness in VQ-VAE for unconditioned samples.